\documentclass[preprint,showpacs,preprintnumbers,amsmath,amssymb,aps,nofootinbib]{revtex4}

\usepackage{graphicx}
\usepackage{dcolumn}

\newcommand{\BE}{\begin{equation}}
\newcommand{\EE}{\end{equation}}
\def\BEA{\begin{eqnarray}}
\def\EEA{\end{eqnarray}}

\begin{document}

\title{QCD Thermodynamics with Three Flavors\\ of Improved Staggered Quarks}

\author{C. Bernard} 
\affiliation{Department of Physics, Washington University, St.~Louis, MO 63130, USA}

\author{T. Burch} 
\affiliation{Institut f\"ur Theoretische Physik, Universit\"at Regensburg, D-93040 Regensburg, Germany}

\author{C. DeTar} 
\affiliation{Physics Department, University of Utah, Salt Lake City, UT 84112, USA}

\author{Steven Gottlieb} 
\affiliation{Department of Physics, Indiana University, Bloomington, IN 47405, USA}

\author{E.B. Gregory} 
\affiliation{Department of Physics, University of Arizona, Tucson, AZ 85721, USA}

\author{U.M. Heller} 
\affiliation{American Physical Society, One Research Road, Box 9000, Ridge, NY 11961-9000}

\author{J. Osborn} 
\affiliation{Physics Department, University of Utah, Salt Lake City, UT 84112, USA}

\author{R. Sugar}
\affiliation{Department of Physics, University of California, Santa Barbara, CA 93106, USA}

\author{D. Toussaint} 
\affiliation{Department of Physics, University of Arizona, Tucson, AZ 85721, USA}

\date{\today}
    
\begin{abstract}
We report on a study of QCD thermodynamics with three flavors of
quarks, using a Symanzik improved gauge action and the Asqtad
$O(a^2)$ improved staggered quark action. Simulations were carried
out with lattice spacings $1/4T$, $1/6T$ and $1/8T$ both for three 
degenerate quarks with masses less than or equal to the strange quark 
mass, $m_s$, and for degenerate up and down quarks with 
masses in the range $0.1 m_s \leq m_{u,d} \leq 0.6 m_s$, and the strange
quark mass fixed near its physical value. We present results for standard
thermodynamics quantities, such as the Polyakov loop, the chiral order 
parameter and its susceptibility. For the quark masses studied to date
we find a rapid crossover rather than a {\it bona fide} phase transition.
We have carried out the first calculations of quark number susceptibilities 
with three flavors of sea quarks. These quantities are of physical interest 
because they are related to event-by-event fluctuations in heavy ion 
collision experiments.  Comparison of susceptibilities at different
lattice spacings show that our results are close to the continuum values.
\end{abstract}
   

\maketitle

\section{INTRODUCTION}

Under ordinary laboratory conditions quarks and gluons, the 
fundamental constituents of quantum chromodynamics (QCD), 
are bound into hadrons. However, as the temperature or
density is increased, the forces among quarks and gluons weaken,
and one expects to find a phase transition or crossover
to a new state of matter, a quark-gluon plasma. The plasma
was a dominant state of matter in the early development of
the universe, and a primary objective of current relativistic heavy ion 
collision experiments is to observe the plasma and determine
its properties.  The behavior of strongly interacting matter
in the vicinity of the phase transition or crossover is 
inherently a strong coupling problem, which at present can 
only be addressed from first principles through lattice gauge
theory calculations.  Among the issues that can uniquely
be addressed by lattice calculations are the nature of the 
transition and the temperature at which it occurs, the properties 
of the plasma, including strange quark content, and the equation 
of state~\cite{REVIEW}. 

In this paper we report on a study of the phase diagram of
high temperature QCD with three flavors of quarks using
improved gauge and quark actions~\cite{PRELIM}. Most lattice 
studies of high temperature QCD have included only the up and down
quarks, but the inclusion of the strange quark is an important
feature of our work~\cite{THREEFLAV}. From chiral models it is expected that 
with two flavors of quarks there is no phase transition at all
for physical values of the up and down quark masses; however,
a strange quark could induce a first order transition, or move
a second order critical point closer to physical quark 
masses~\cite{CHIRALMODELS}. Furthermore, the production of
strangeness is expected to be an important signal for the
plasma in heavy ion experiments. 

We are considering two cases in our study:
1) all three quarks have the same mass $m_q$; and 2) the two lightest
mass quarks have equal mass $m_{u,d}$, while the mass of the third
quark is fixed at approximately that of the strange quark $m_s$. We 
refer to these cases as $N_f=3$ and $N_f=2+1$, respectively. 
For $N_f=3$ we have worked with quark masses in the range
$0.2\, m_s\leq m_q \leq	m_s$, while for $N_f=2+1$ we have carried
out simulations with $0.1\, m_s \leq m_{u,d} \leq 0.6\, m_s$.
We have monitored standard thermodynamic quantities such as
the plaquette, the Polyakov loop, the chiral order parameter and
its susceptibility. For the quark masses we have studied to date, we find 
rapid crossovers, which sharpen as the quark mass is reduced, rather 
than a {\it bona fide} phase transition. This is in agreement with earlier work~\cite{REVIEW}.
We have also measured the quark number susceptibilities, 
which provide excellent signals for the crossover, and which are directly
related to event-by-event fluctuations in heavy ion collisions~\cite{EbyE}. 
The data indicate that our results for these susceptibilities are
quite close to their continuum values.

\section{THE SIMULATIONS}

\begin{figure}
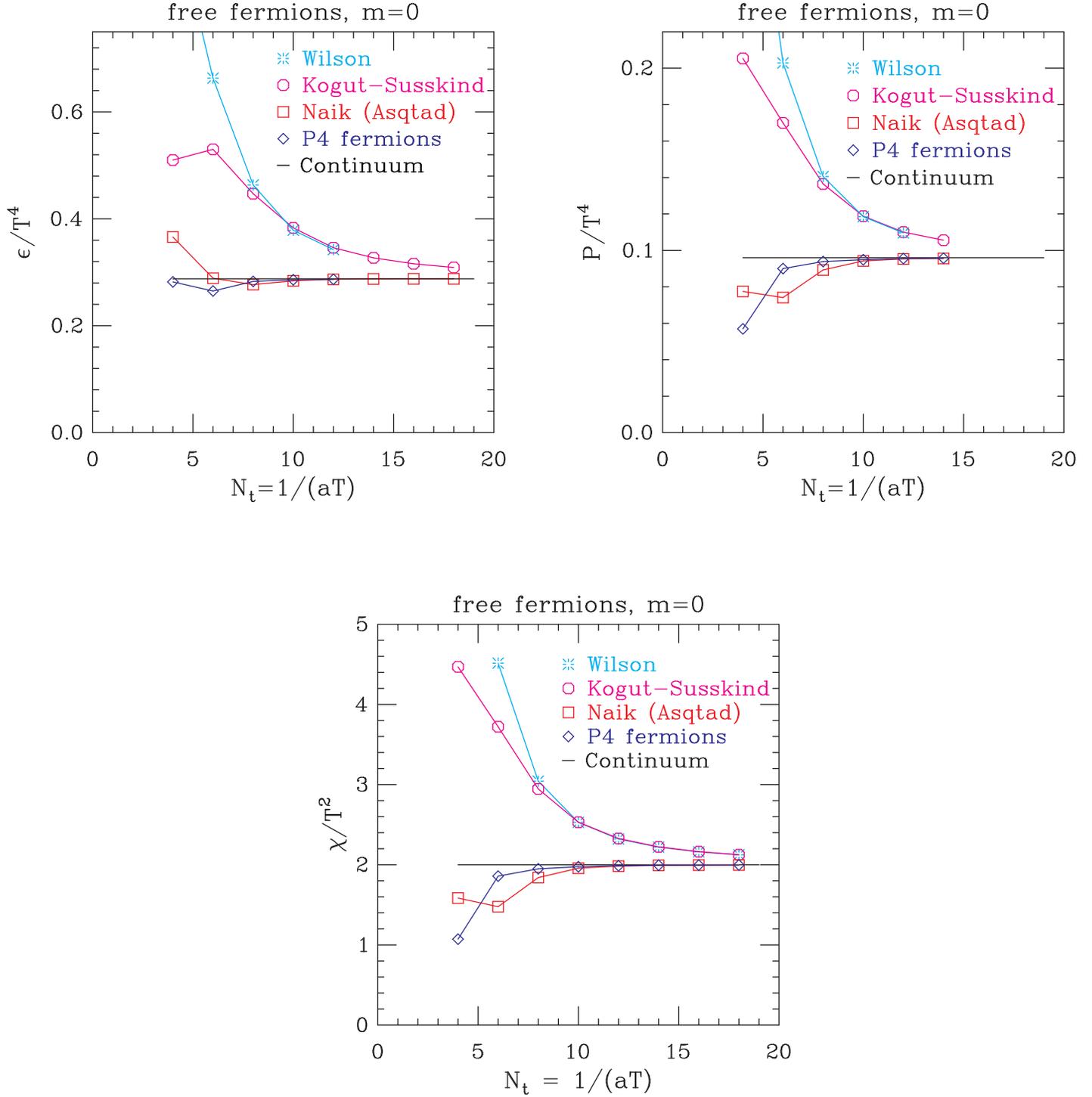

\centerline{{\includegraphics[width=3.5in]{f_energy.ps}}\hspace{1.0cm}
{\includegraphics[width=3.5in]{f_pressure.ps}}}
\vspace{1.5cm}
\centerline{\includegraphics[width=3.5in]{qno_free2.ps}}
\caption{The energy density, pressure and quark number susceptibility
for free, massless quarks as a function of the temporal lattice size, 
$N_t$, for conventional and improved actions. For free fermions the Asqtad 
action reduces to the Naik action. The $P4$ action is the improved staggered 
fermion action studied by the Bielefeld group~\cite{BIELEFELD}.
\label{fig:EPQFIG}
}
\end{figure}

Our simulations are carried out with a one-loop Symanzik improved gauge
action~\cite{GA}  and the Asqtad quark action~\cite{ASQTAD}.
Both the gauge and quark actions have all lattice artifacts removed 
through order $a^2$ at tree level, where $a$ is the lattice spacing, 
and are tadpole improved.  Thus, the leading order finite lattice 
spacing artifacts are of order $a^2\alpha_s$ and $a^4$. We and others
have used this action over the last few years to perform a wide variety of
zero-temperature calculations~\cite{SGLAT03}. It has a number of features
that make it particularly well suited for high temperature studies.
The Asqtad action has considerably better dispersion relations for free 
quarks than the standard Kogut-Susskind and Wilson actions~\cite{SPECPAPER}, 
which markedly decreases lattice artifacts above the transition. This
is also true of the P4 action used by the Bielefeld group in its studies
of high temperature QCD~\cite{BIELEFELD}. The improvement in dispersion
relations is illustrated in Fig.~\ref{fig:EPQFIG}, where we plot the energy, 
pressure and quark number susceptibility for free massless quarks as a 
function of lattice spacing.  
The Asqtad action exhibits excellent scaling properties in the lattice 
spacing~\cite{IMP_SCALING}, which accelerates the approach to the 
continuum limit. Finally, taste symmetry breaking is much smaller for 
the Asqtad action, than for the conventional Kogut-Susskind action. 
Our spectrum studies with the Asqtad action indicate that for 
lattices with eight to ten time slices, the kaon is heavier than 
the heaviest non-Goldstone pion in the neighborhood of the finite 
temperature transition or crossover, a condition which requires
much smaller lattice spacings with the conventional Kogut-Susskind
action. It is, of course, difficult to study the effects
of the strange quark on the transition if this condition is not
fulfilled.

We have used the refreshed hybrid molecular dynamics R~algorithm~\cite{RALG}
to generate gauge configurations. In integrating the molecular dynamics
equations of motion we used a time step $dt$ equal to the smaller
of 0.02 and $\frac{2}{3}m_{u,d}$. The momenta conjugate to the gauge
fields were refreshed every molecular dynamics time unit, which consisted
of $1/dt$ time steps. In the vicinity of the crossover we generated
2,000 equilibrated molecular dynamics time units at each value of the 
gauge coupling and quark masses we studied. The runs well above and
below the crossover were shorter. Measurements of standard thermodynamics 
quantities were made after each time unit, and gauge configurations were 
saved every five time units for separate measurements of the quark 
number and $\bar\psi\psi$ susceptibilities. Ten random sources were
used in the measurements of $\bar\psi\psi$ made at the end of each
time unit, and 100 random sources were used in the calculations of
the susceptibilities. The values of the chiral order parameter obtained
in these two measurements agreed to within statistics.

We have attempted to vary the temperature while keeping all other physical
quantities constant. To this end, for the $N_f=3$ study we have performed 
a set of spectrum calculations at lattice spacings $a=0.125$~fm and 0.18~fm
with $m_q=m_s$, $0.6\, m_s$, $0.4\, m_s$, and $0.2\, m_s$. 
We determine the lattice spacing from the static $\bar Q Q$ potential,
and express it and other dimensionful quantities in terms of $r_1$ defined by
$r_1^2\,F_{\bar Q Q}(r_1)=1$. Using results from the 1S-2S and 1P-1S
splitting in the $\Upsilon$ spectrum~\cite{UPSILON}, we have found
$r_1=0.317\pm 0.007$~fm in the continuum and chiral limits~\cite{NEWSPEC}. 
We determine $m_s$ from the 
requirement that $m_{\eta_{ss}}/m_\phi \approx 0.673$, where $\eta_{ss}$
is an ``unmixed'' pseudoscalar meson made of an $s$ and $\bar s$ quark.
We would like to carry out our thermodynamics studies with three equal 
mass quarks for $m_{\eta_{ss}}/m_\phi$
fixed, but this quantity will, of course, vary slightly with lattice spacing
if we keep $m_q/m_s$ fixed. So, at $a=0.18$~fm we perform linear
interpolations of $m_{\eta_{ss}}^2$ and $m_\phi$ in the quark mass 
to determine
the precise values of $m_q$ for which $m_{\eta_{ss}}/m_\phi$ will take on the
values found at $a=0.125$~fm. Then to determine the values of $am_q$, $a$
and $T=1/aN_t$ for thermodynamics studies with other values of $a$,
we perform interpolations or extrapolations with
a form due to Allton, inspired by asymptotic freedom~\cite{ALTON}. 
We first define

\begin{equation}
f(g^2) = (b_0g^2)^{-b_1/(2b^2_0)}\; e^{ - 1/2b_0 g^2} ,
\end{equation}

\noindent
where $b_0$ and $b_1$ are the universal beta-function coefficients
for massless three-flavor QCD, and $g^2$ is the bare lattice
coupling. $f(g^2)$ is basically $a/\Lambda_L$. We then determine $a$
as a function of $g^2$ from the interpolation formula

\begin{equation}
a(g^2)/r_1 = c_0\, f(g^2) \left[ 1 +  c_2 g^2 f^2(g^2) \right] .
\label{eq:interpa}
\end{equation}

\noindent
The second term in Eq.\,(\ref{eq:interpa}) is an $O(g^2\,a^2)$
correction to the asymptotic freedom formula. The coefficients
$c_0$ and $c_2$ are determined from the measured values of
$a/r_1$ at $a=0.18$ and 0.125~fm. A similar interpolation formula
is used to determine the lattice quark mass along the line
of fixed $m_{\eta_{ss}}/m_\phi$,

\begin{equation}
am_q(g^2) = d_0\, (b_0 g^2)^{-4/9} f(g^2) \left[ 1 +  d_2 g^2 f^2(g^2) \right].
\label{eq:interpm}
\end{equation}
Here we have included the anomalous dimension of the mass.
In the initial stages of our work we simply made linear interpolations
of $\ln(a)$ and $\ln(am_q)$ in $1/g^2$ between the anchor points at $a=0.18$
and 0.125~fm, which gave results in good agreement with those of 
Eqs.\,(\ref{eq:interpa}) and (\ref{eq:interpm}). 
However, use of the more sophisticated formulae became
critical for points outside the anchors.

Our approach for thermodynamics studies with $m_{u,d}< m_s$ is quite similar.
In this case we wish to vary the temperature keeping both $m_{\pi}/m_{\rho}$
and $m_{\eta_{ss}}/m_\phi$ fixed. We have carried out spectrum
studies at $a=0.125$ and 0.18~fm for light quark masses $m_{u,d} = m_s$, 
$0.6\, m_s$, $0.4\, m_s$, $0.2\, m_s$, and $0.1\, m_s$, and at $a=0.09$~fm with
$m_{u,d}=m_s$, $0.4\, m_s$ $0.2\, m_s$ and $0.1\, m_s$ all with the mass of 
the heavy quark fixed close to $m_s$~\cite{SPECPAPER, NEWSPEC}. 
The strange quark mass  was determined from the
spectrum calculations with three equal mass quarks.  In our spectrum runs at
$a=0.125$~fm~\cite{SPECPAPER} we found that $m_{\eta_{ss}}$ varied by less than
2\% and $m_\phi$ by less than 1\% for $0.2\, m_s \leq m_{u,d}\leq m_s$,
and the heavy quark mass held fixed at $m_s$. So, the neglect of the
dependence of $m_{\eta_{ss}}$ and $m_\phi$ on $m_{u,d}$ is well justified.
In these studies we performed linear interpolations of $m_{\pi}^2$ and 
$m_{\rho}$ at $a=0.18$~fm to determine the values of $m_{u,d}$ for which 
$m_\pi/m_\rho$ takes on the values found at $a=0.125$~fm. Then, for other 
values of $a$ we interpolate or extrapolate $a/r_1$ using 
Eq.\,(\ref{eq:interpa}), 
and $am_{u,d}$ and $am_s$ using Eq.\,(\ref{eq:interpm}). We again determine the 
values of $c_0$, $c_2$, $d_0$ and $d_2$ from the spectrum runs at 
$a=0.125$~fm and 0.18~fm. For $m_{u,d}=m_s$, $0.4\, m_s$, $0.2\, m_s$
and $0.1\, m_s$, where we have spectrum data at $a=0.09 $~fm, 
we can add  $f^4(g^2)$ terms 
to the right hand sides of Eqs.\,(\ref{eq:interpa})
and~(\ref{eq:interpm}). These added terms make only a few percent difference
in the lattice spacing and quark masses extrapolated to $10\, T_C$,
giving us confidence in the interpolations and extrapolations
used in our study. We have recently combined data from all
zero temperature runs we have made to date to obtain a smooth interpolation
formula of $\ln(r_1/a)$ as a function of quark mass and gauge coupling~\cite{NEWSPEC}. 
The results are in excellent agreement with those obtained from Eq.\,(2).

\section{RESULTS FOR $\bf N_f=3$}

For three equal mass quarks, $N_f=3$, we have carried out thermodynamics 
studies on lattices with four, six and eight time slices, and aspect 
ratio $N_s/N_t=2$. Here $N_s$ and $N_t$ are the spatial and temporal 
dimensions of the lattice in units of the lattice spacing.  We also performed 
simulations with aspect ratio three for $N_t=4$, and obtained results
that are indistinguishable from those with aspect ratio two. The spectrum 
calculations and interpolations described above allowed us to determine 
the values of the quark mass, $m_q$, that keep $m_{\eta_{ss}}/m_\phi$ 
fixed as the gauge coupling is varied. They also enabled us to determine 
the value of the lattice spacing, and therefore the temperature for each run.

\begin{figure}
\centerline{\includegraphics[width=6.0in]{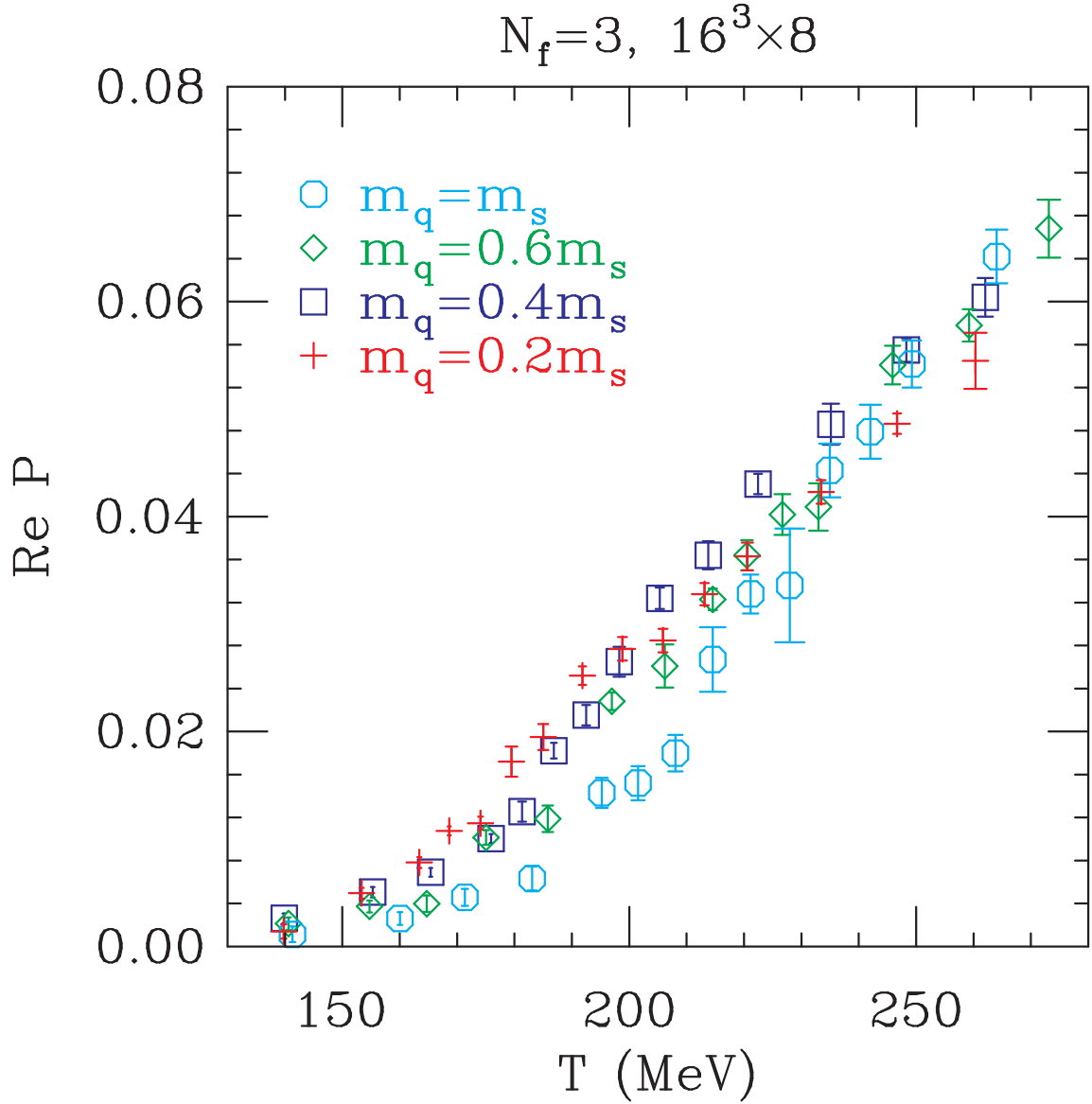}}
\caption{The real part of the Polyakov loop as a function of temperature
on $16^3\times 8$ lattices for
three degenerate flavors of quarks with masses $m_q/m_s=1.0$, 0.6, 0.4 and 0.2.
\label{fig:rp_nf3_nt8}
}
\end{figure}

In Fig.~\ref{fig:rp_nf3_nt8} we plot the real part of the Polyakov loop as
a function of temperature on $16^3\times 8$ lattices. In this figure, as
elsewhere, we give the values of $m_q/m_s$ for $a=0.125$~fm. The corresponding
values of $m_{\eta_{ss}}/m_\phi$ are given in Table~\ref{tab:massratio}. 
The Polyakov loop shows a crossover from confined behavior at low temperature 
to deconfined behavior at high temperature. There is a slight trend for the
temperature dependence of the Polyakov loop to be steeper for larger quark
masses. This is to be expected, since at sufficiently large quark masses,
it is a {\it bona fide} order parameter.  The insensitivity of the Polyakov 
loop to the quark mass at fixed temperature or lattice spacing is,
perhaps, not surprising. We have determined the lattice spacing from the heavy
quark potential, and in our spectrum runs we adjusted the coupling constant to
keep the lattice spacing fixed as the quark mass is varied. Since the Polyakov
loop, like the heavy quark potential, is determined from measurements of
purely gluonic operators, our procedure is likely to minimize the dependence
of the Polyakov loop on the quark mass.

\vspace{8mm}

\begin{table}[ht]
\begin{center}
\begin{tabular}{|c|c|c|c|}
\hline
\multicolumn{2}{|c|}{\rule[-3mm]{0mm}{8mm}$N_f=3$}& \multicolumn{2}{|c|}{\rule[-3mm]{0mm}{8mm}$N_f=2+1$}\\
\hline\hline
\rule[-3mm]{0mm}{8mm}$m_q/m_s$ & $m_{\eta_{ss}}/m_\phi$ &  $m_{u,d}/m_s$ & $m_\pi/m_\rho$  \\
\hline
\rule[0mm]{0mm}{6mm}1.0 & 0.673   & 1.0  & 0.673    \\
\rule[0mm]{0mm}{0mm}0.6  & 0.583  & 0.6  & 0.582   \\
\rule[0mm]{0mm}{0mm}0.4  & 0.504  & 0.4  & 0.509   \\
\rule[0mm]{0mm}{0mm}0.2  & 0.404  & 0.2  & 0.392   \\
\rule[-2mm]{0mm}{2mm}     &        & 0.1  & 0.298   \\
\hline
\end{tabular}
\vspace{4mm}
\caption{In the first column we show the value of $m_q/m_s$ at lattice spacing
0.125~fm, which produced the $m_{\eta_{ss}}/m_\phi$ ratio shown in the second 
column for spectrum calculations with three equal mass quarks. In the third 
column we give the value of $m_{u,d}/m_s$ which produced the $m_\pi/m_\rho$ 
ratio shown in the fourth column for spectrum calculations with two equal mass
light quarks, and the mass of the heavy quark fixed at $m_s$.
\label{tab:massratio}
}
\end{center}
\end{table}

\begin{figure}
\centerline{\includegraphics[width=6.0in]{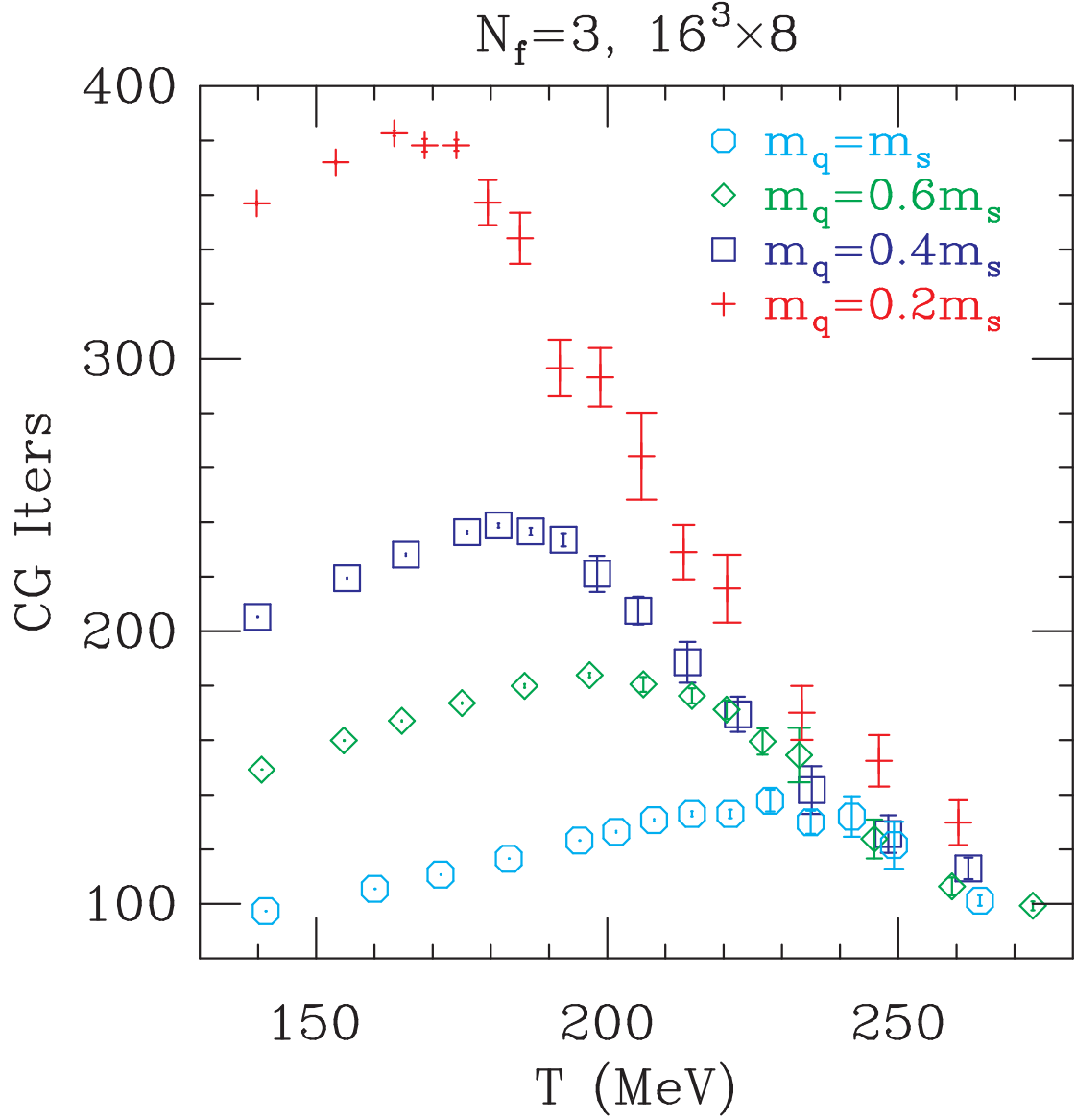}}
\caption{The number of conjugate gradient iterations required for
convergence of the inversion of the Dirac operator for
three degenerate flavors of quarks with masses $m_q/m_s=1.0$, 0.6, 0.4 and 0.2
on $16^3\times 8$ lattices.
\label{fig:cg_nf3_nt8}
}
\end{figure}

\begin{figure}
\centerline{\includegraphics[width=6.0in]{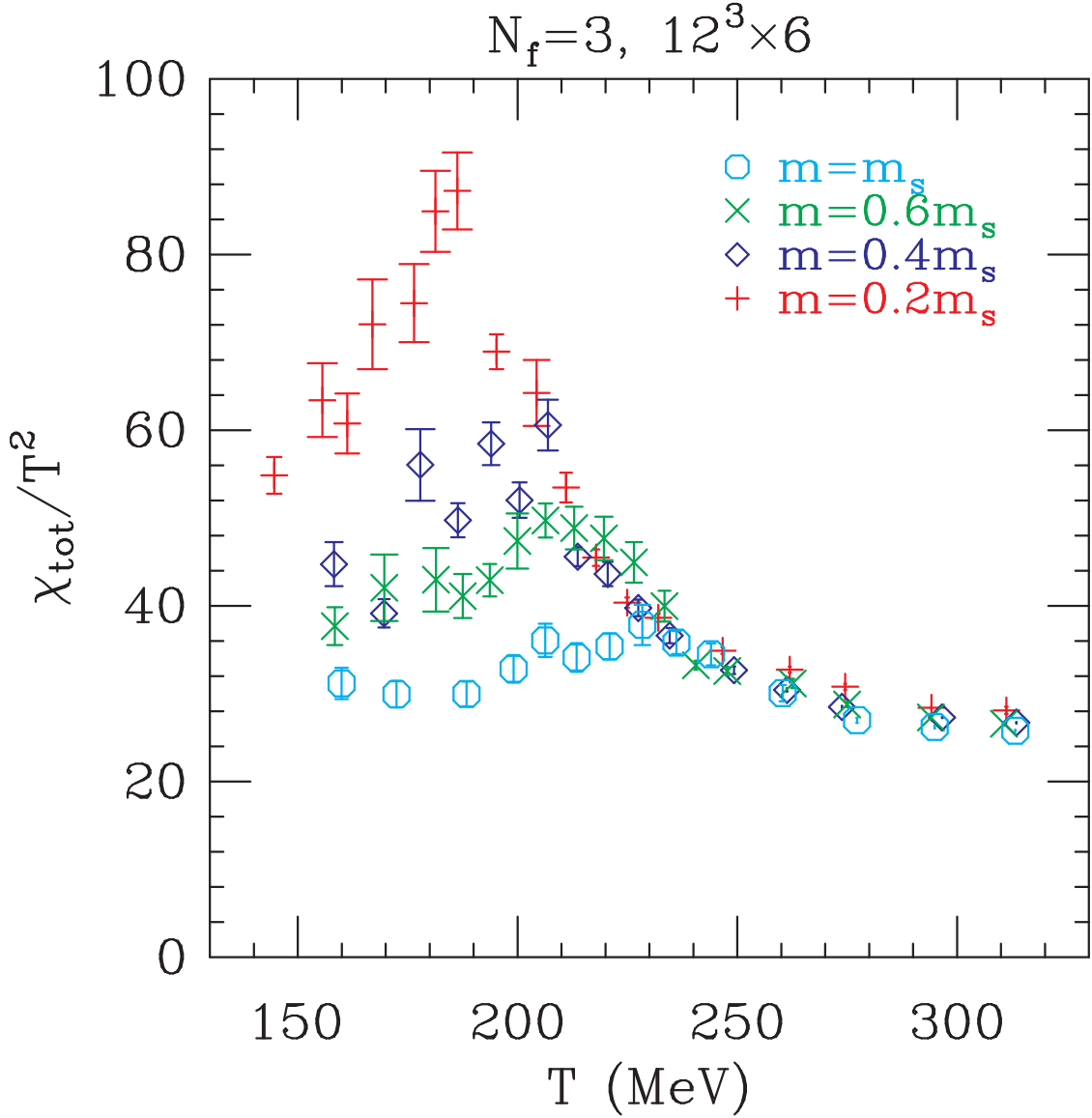}}
\caption{The $\bar\psi\psi$ susceptibility as a function of temperature
for three equal mass quarks on
$12^3\times 6$ lattices. Results are shown for quark masses 
$m_q/m_s=1.0$, 0.6, 0.4, and 0.2.
Note the increase in the height of the peak as the quark mass is decreased.
\label{fig:chi_tot_nf3_nt6}
}
\end{figure}

The number of conjugate gradient iterations varies as the temperature is
changed reflecting a changing condition number for the Dirac operator
as the spectrum of states changes.  We show this quantity in 
Fig.~\ref{fig:cg_nf3_nt8} for the four values of the quark mass we have 
studied on $16^3\times 8$ lattices. The sharpening of the crossover as the 
quark mass is reduced is evident. The $\bar\psi\psi$ susceptibility provides 
a more physical signal for the crossover. It is given by

\begin{equation}
\chi_{\rm tot} = \frac{\partial}{\partial m} \langle\bar \psi \psi\rangle
 = \frac{T}{V_s} \langle {\rm Tr}\; M^{-1} {\rm Tr}\; M^{-1} \rangle - \frac{T}{V_s} \langle
{\rm Tr}\; M^{-2}\rangle - \frac{T}{V_s} \langle {\rm Tr}\; M^{-1} \rangle^2 ~,
\label{eq:chi_tot}
\end{equation}

\noindent
where $M$ is the fermion matrix, $T$ the temperature, and $V_s$ the spatial volume. 
The traces in Eq.\,(\ref{eq:chi_tot}) were, as usual,  
evaluated using the identity

\begin{equation}
{\rm Tr}\; O = \langle R^*\, O\,  R \rangle_R\; ,
\end{equation} 

\noindent
where $\langle\rangle_R$ indicates an average over vectors, $R$, of Gaussian
random numbers. We used 100 vectors of
random numbers for each gauge configuration. Of course, care must be taken 
not to use the same vector of random numbers in evaluating the product 
of traces in the first term of Eq.\,(\ref{eq:chi_tot}). 
We plot $\chi_{\rm tot}$ 
in Fig.~\ref{fig:chi_tot_nf3_nt6} for quark masses $m_s$, $0.6\, m_s$, 
$0.4\, m_s$, and $0.2\, m_s$ on $12^3\times 6$ lattices. Note the 
increase in the height of the peak as the quark mass is decreased.

\begin{figure}
\centerline{\includegraphics[width=6.0in]{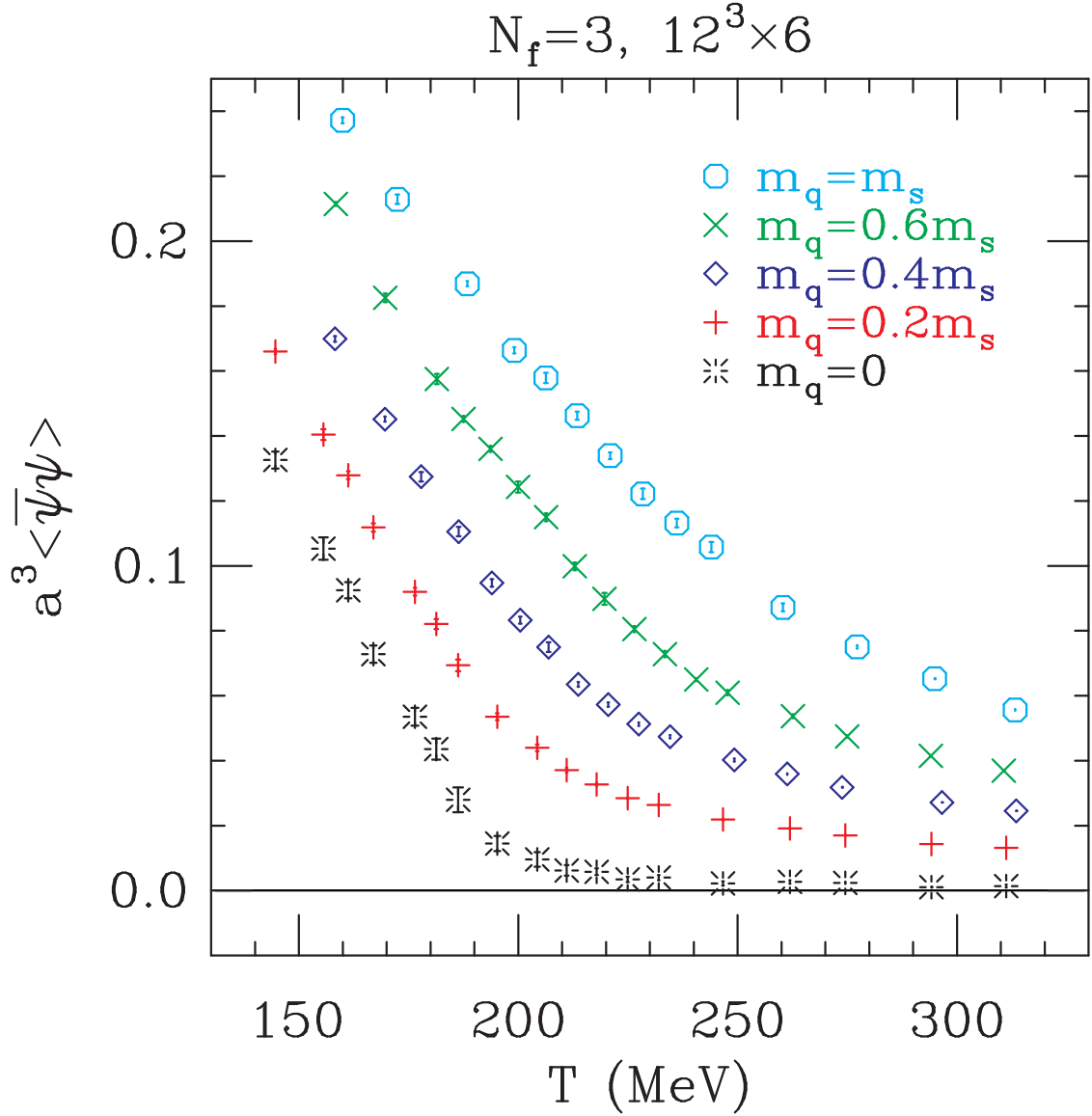}}
\caption{The chiral order parameter, $\langle\bar\psi\psi\rangle$, 
as a function of temperature on
$12^3\times 6$ lattices for $N_f=3$. The bursts are linear extrapolations
of $\langle\bar\psi\psi\rangle$ for the two lowest quark masses
to $m_q=0$ at fixed temperature.
\label{fig:pbp_nf3_nt6}
}
\end{figure}

\begin{figure}
\centerline{\includegraphics[width=6.0in]{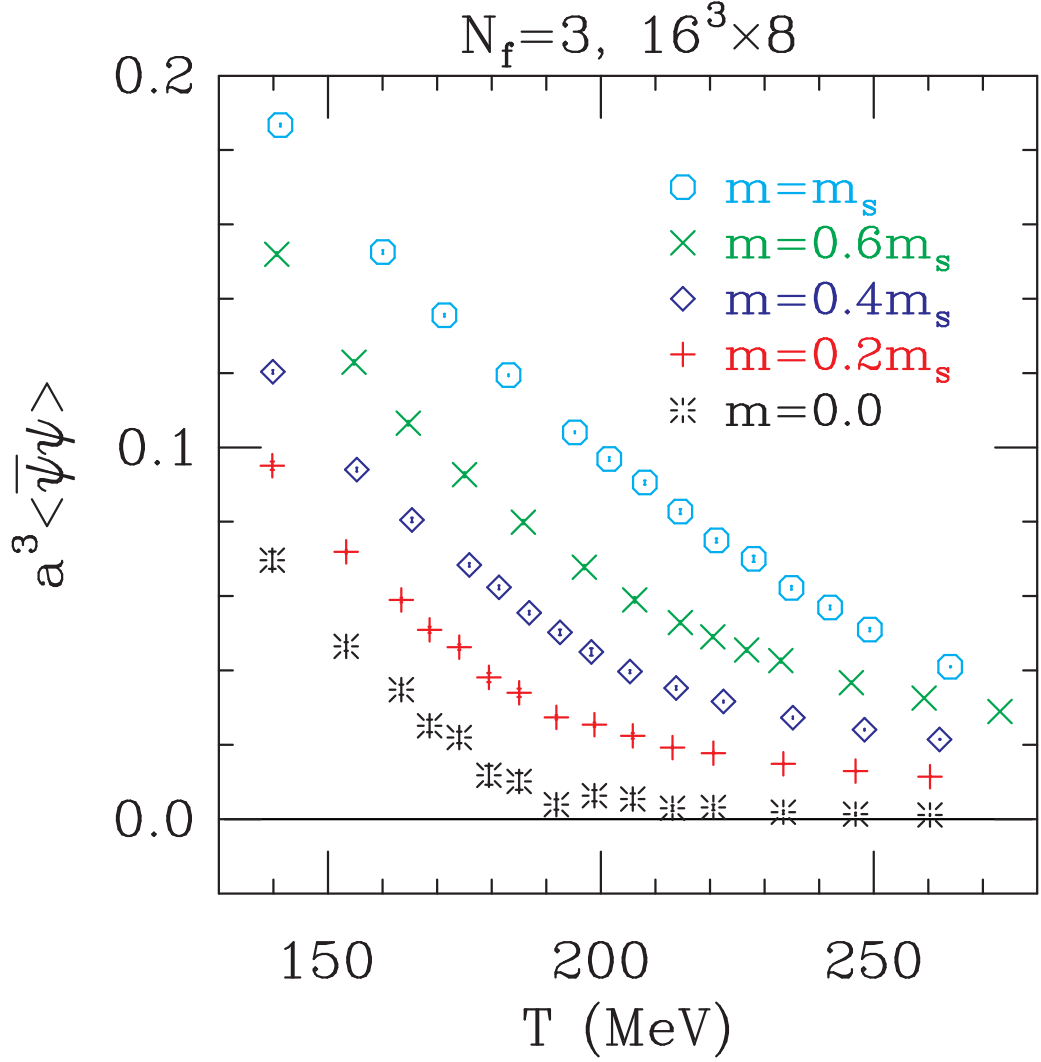}}
\caption{The chiral order parameter $\langle\bar\psi\psi\rangle$ 
as a function of temperature on 
$16^3\times 8$ lattices for $N_f=3$. The bursts are linear extrapolations 
of $\langle\bar\psi\psi\rangle$ for the two lowest quark masses
to $m_q=0$ at fixed temperature.
\label{fig:pbp_nf3_nt8}
}
\end{figure}

In Figs.~\ref{fig:pbp_nf3_nt6} and~\ref{fig:pbp_nf3_nt8} we show the
chiral order parameter, $\bar\psi\psi$ as a function of temperature for
$m_q=m_s$, $0.6\,m_s$, $0.4\, m_s$, and $0.2\, m_s$  on $12^3\times 6$ 
and $16^3\times 8$ lattices, respectively. The bursts in these figures 
are linear extrapolations of $\bar\psi\psi$ in the quark mass to $m_q=0$ for 
fixed temperature.  These figures suggest that for $m_q=0$ there is unlikely 
to be a phase transition for temperatures above 190~MeV, but one could occur 
at or below that value. For sufficiently high
temperatures and small quark masses,
one expects a linear extrapolation in the quark mass to be valid, so the
vanishing of $\bar\psi\psi$ at $m_q=0$  indicates that the system is in the
chiral symmetric phase. However, a non-vanishing value of the extrapolated
$\bar\psi\psi$ could indicate a first order transition at a non-zero quark mass,
or simply a breakdown in the validity of the linear extrapolation in the quark mass.

\section{RESULTS FOR $N_f=2+1$}

The $N_f=2+1$ thermodynamics studies were carried out primarily on $8^3\times 4$,
$12^3\times 6$ and $16^3\times 8$ lattices. In addition, a number of runs 
were made on $18^3\times 6$ lattices to check for finite spatial size effects, which 
turned out to be negligible. In this phase of our work, we performed simulations
with two degenerate light quarks and the heavy quark mass held approximately
equal to that of the strange quark.  The spectrum calculations and 
interpolations described above allowed us to determine curves of constant 
physics in the coupling constant-quark mass plane, and to determine the 
lattice spacing, and therefore the temperature for each run. 

\begin{figure}
\centerline{\includegraphics[width=6.0in]{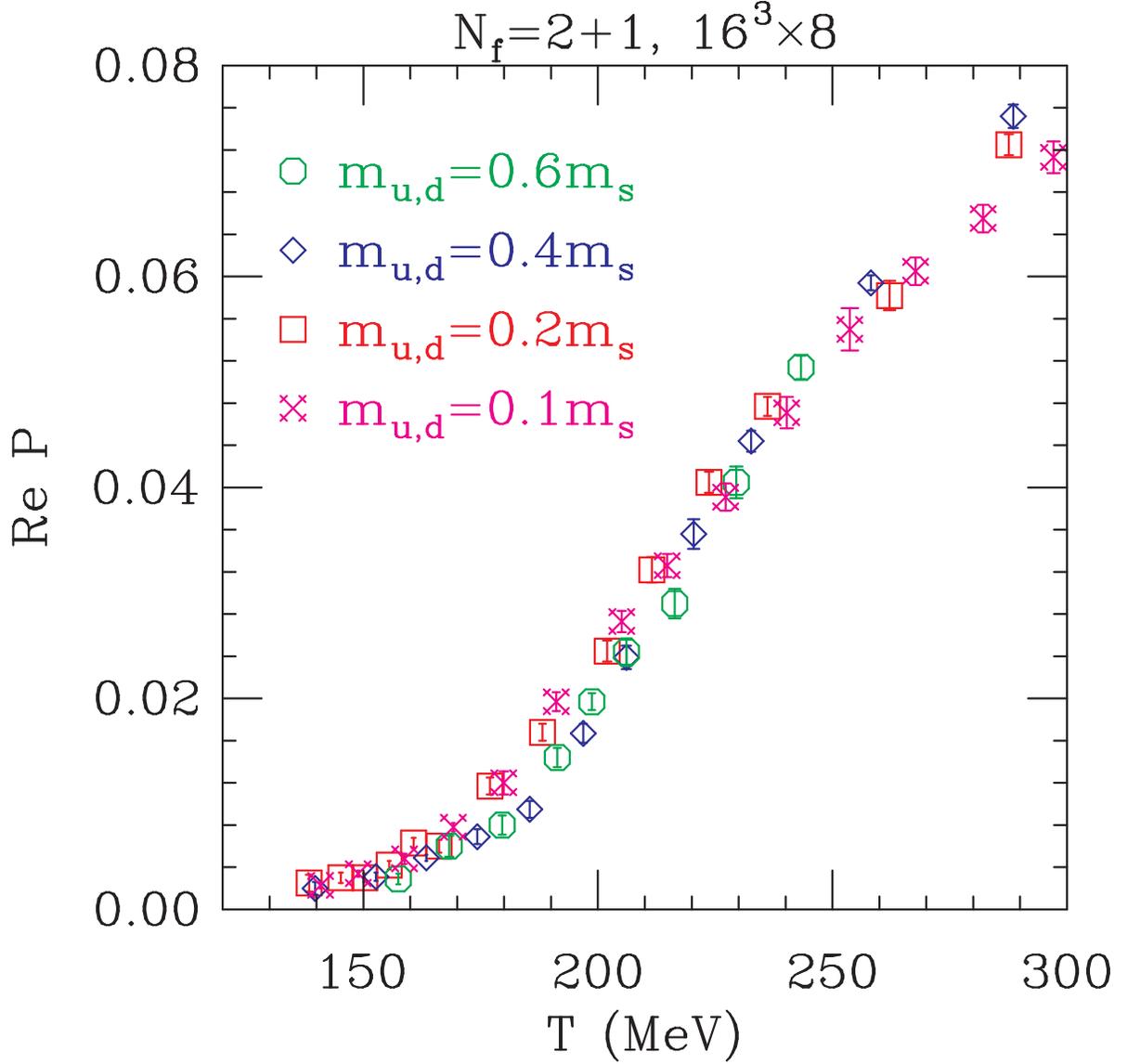}}
\caption{The real part of the Polyakov loop as a function of temperature
for two light and one heavy quark on $16^3\times 8$ lattices. Results are shown
for light quark masses $m_{u,d}/m_s=0.6$, 0.4, 0.2 and 0.1. The
mass of the heavy quark is fixed at $m_s$.
\label{fig:rp_nf21_nt8}
}
\end{figure}

\begin{figure}
\centerline{\includegraphics[width=6.0in]{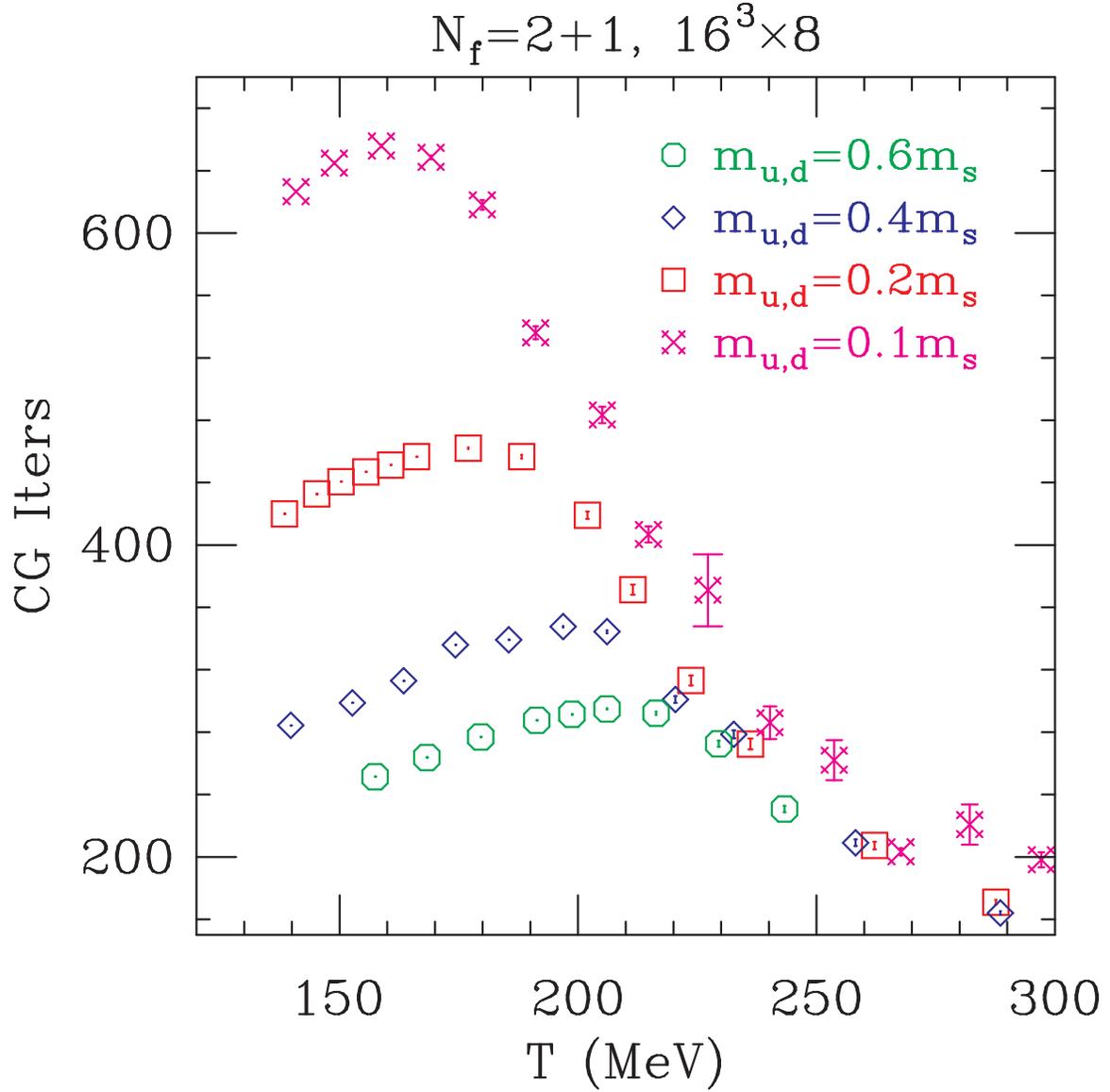}}
\caption{The number of conjugate gradient iterations required for convergence
of the inversion of the Dirac operator of light quarks
on $16^3\times 8$ lattices. Results are shown for light quark masses
$m_{u,d}/m_s=$, 0.4, 0.2 and 0.1. The mass of the heavy quark
is fixed at $m_s$.
\label{fig:cg_nf21_nt8}
}
\end{figure}

\begin{figure}
\centerline{\includegraphics[width=6.0in]{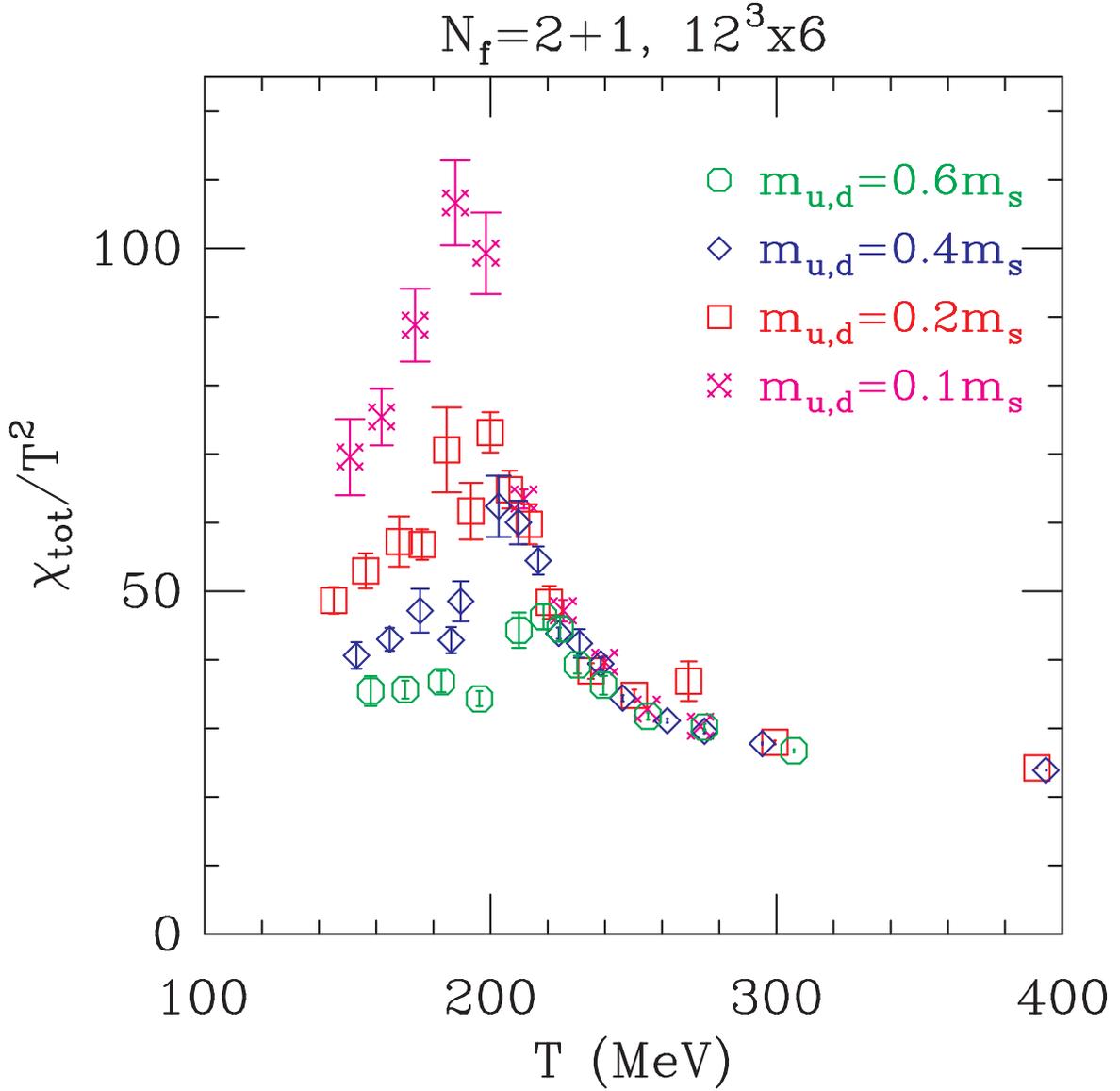}}
\caption{The $\bar\psi\psi$ susceptibility as a function of temperature
for two light and one heavy quark on
$12^3\times 6$ lattices. Results are shown
for light quark masses $m_{u,d}/m_s=0.6$, 0.4, 0.2 and 0.1. The
mass of the heavy quark is fixed at $m_s$.  
\label{fig:chi_tot_nf21_nt6}
}
\end{figure}

In Fig.~\ref{fig:rp_nf21_nt8} we plot the real part of the Polyakov loop
as a function of temperature on $16^3\times 8$ lattices for the four values
of $m_{u,d}$ studied to date. As in the $N_f=3$ study, we observe a crossover
from confined to deconfined behavior, rather than a sharp transition, and
little dependence on the light quark mass.  In Fig.~\ref{fig:cg_nf21_nt8}
we show the number of conjugate gradient iterations needed for inversion of the
Dirac operator for the light quarks on $16^3\times 8$ lattices, and in
Fig.~\ref{fig:chi_tot_nf21_nt6} we plot the $\bar\psi\psi$ susceptibility
for the up and down quarks 
on $12^3\times 6$ lattices. In both figures the sharpening of the crossover
as the light quark masses are reduced is evident.

\begin{figure}
\centerline{\includegraphics[width=6.0in]{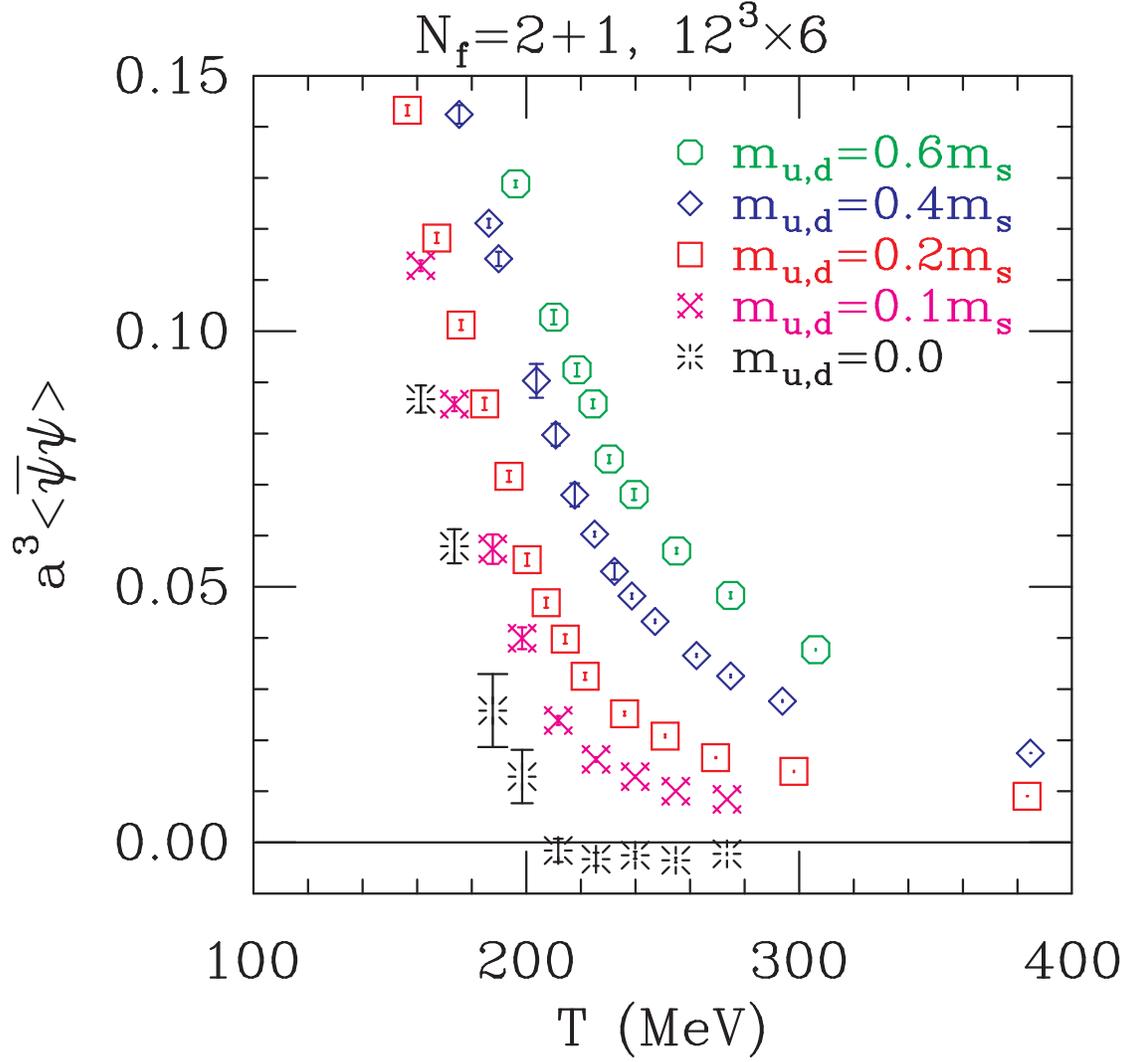}}
\caption{The chiral order parameter $\bar\psi\psi$ as a function of temperature
on $12^3\times 6$ lattices. The bursts are linear extrapolations in the quark 
mass to $m_{ud}=0$ for fixed temperature.
\label{fig:pbp_nf21_nt6}
}
\end{figure}

\begin{figure}
\centerline{\includegraphics[width=6.0in]{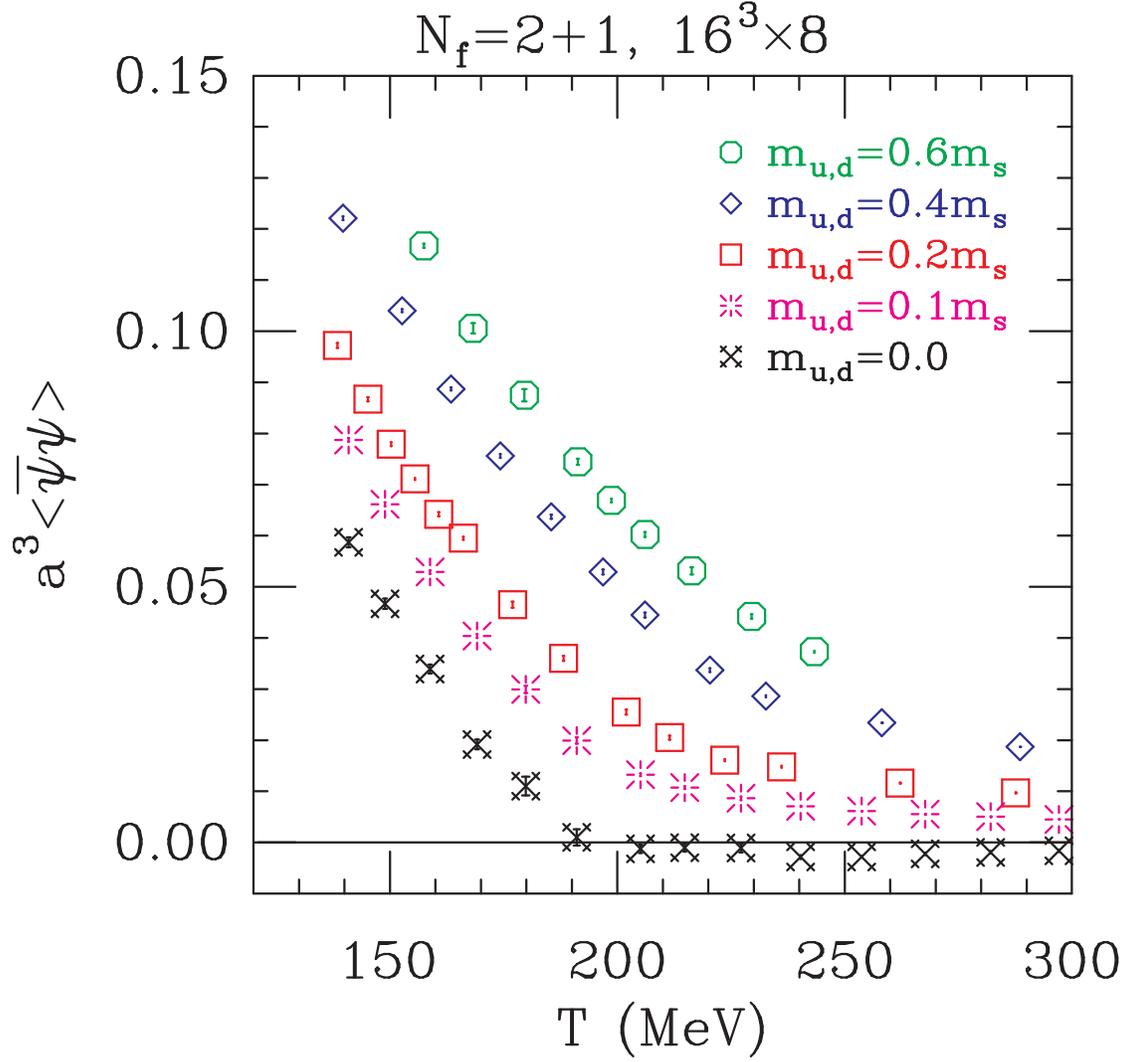}}
\caption{The chiral order parameter $\bar\psi\psi$ on $16^3\times 8$ lattices.
The bursts are linear extrapolations in the quark mass to $m_{ud}=0$
for fixed temperature.
\label{fig:pbp_nf21_nt8}
}
\end{figure}

In Figs.~\ref{fig:pbp_nf21_nt6} and~\ref{fig:pbp_nf21_nt8} we plot the 
chiral order parameter as a function of temperature on $12^3\times 6$ 
and $16^3\times 8$ lattices.  The octagons, diamonds, squares and fancy crosses
are data for $m_{u,d}=0.6\, m_s$, $0.4\, m_s$,  $0.2\, m_s$ and $0.1\, m_s$
respectively, and the bursts are linear extrapolations in $m_{u,d}$ 
for fixed temperatures using the $0.2\, m_s$ and $0.1\, m_s$ points.
The fact that the linear extrapolation of the chiral order parameter
becomes slightly negative suggests that in this region of temperature and quark 
mass $\bar\psi\psi$ is nonlinear, as would be expected if there were
a {\it bona fide} critical point for $m_{u,d}\ge 0$.

\section{QUARK NUMBER SUSCEPTIBILITIES}

In order to study the quark number 
susceptibilities~\cite{SUSC,GG,KAR,REB,CMT,AV}, we introduce
chemical potentials $\mu_\alpha$ coupled to a set of mutually
commuting conserved charges $Q_\alpha$. The partition function
can then be written

\begin{equation}
Z = {\rm exp}\left[-\beta(H-\sum_\alpha\, \mu_\alpha Q_\alpha)\right].
\end{equation}
\noindent
The quark number susceptibilities can be related to event-by-event 
fluctuations in heavy ion collisions~\cite{EbyE} by the 
fluctuation-dissipation theorem

\begin{equation}
\chi_{\alpha,\beta}(T) =
\left\langle(Q_\alpha-\langle Q_\alpha\rangle) (Q_\beta-\langle Q_\beta\rangle)  \right\rangle 
\propto \frac{T}{V_s}\, \frac{\partial^2 \log Z}{\partial{\mu_\alpha}\partial{\mu_\beta}}\, , 
\label{eq:chi}
\end{equation}

\noindent
We work at $\mu_\alpha=0$, so the brackets, $\langle\rangle$, in
Eq.~(\ref{eq:chi}) indicate averages weighted by the standard, real
Euclidean action for QCD, and $\langle Q_\alpha\rangle=0$. 
The charges we consider are bilinears in the
quark fields, so it is possible to write $\chi_{\alpha,\beta}$ in
the form

\begin{equation}
\chi_{\alpha,\beta} = C_{\alpha,\beta} + D_{\alpha,\beta}
\end{equation}

\noindent
where $C_{\alpha,\beta}$ is the contribution to $\chi_{\alpha,\beta}$
of connected diagrams in which quark fields from charge $Q_\alpha$ 
contract with those from $Q_\beta$, whereas $D_{\alpha,\beta}$
is the contribution of the disconnected diagrams in which quark fields
within each charge contract.

If we take the charges to be the number operators for up, down and
strange quarks, then because we have set $m_u=m_d$, $\chi$ takes
the form

\begin{equation}
\chi = \left(
\begin{array}{ccc}
C_{l,l} + D_{l,l} & D_{l,l}             & D_{l,s}\\
D_{l,l}           & C_{l,l} + D_{l,l}   & D_{l,s}\\
D_{l,s}           & D_{l,s}             & C_{s,s} + D_{s,s}
\end{array}
\right),
\end{equation}
where the rows and columns of the matrix are labeled by $u$, $d$
and $s$ in that order, and $l$ stands for the equal up and down
quark matrix elements.

It appears more physical to take the three independent charges
to be the z-component of isospin, $Q_I$, the
hypercharge, $Q_Y$, and the baryon number, $Q_B$, where in the continuum

\begin{eqnarray}
Q_I &=& \frac{1}{2}\,\int d^3x \, \Psi^\dagger (x)\lambda_3 \Psi(x) \\
Q_Y &=& \frac{1}{\sqrt{3}}\,\int d^3x \, \Psi^\dagger (x)\lambda_8 \Psi(x) \\
Q_B &=& \frac{1}{3}\int d^3x \, \Psi^\dagger (x)\cdot \Psi(x). 
\end{eqnarray}

Here $\Psi(x)$ is a three-component column vector, whose components are
the up, down and strange quark fields, and $\lambda_3$ and $\lambda_8$ are
the standard diagonal generators of SU(3) in the fundamental representation,
$\lambda_3={\rm diag}(1,-1,0)$ and 
$\lambda_8={\rm diag}(1,1,-2)/\sqrt{3}$.  Using 
these charges, $\chi$ can be written in the form

\begin{equation}
\chi = \left(
\begin{array}{ccc}
\frac{1}{2}C_{l,l}        &       0     &    0      \\
0 & \frac{2}{9}(C_{l,l}+2C_{s,s}) 
+\frac{4}{9}(D_{l,l}-2D_{l,s}+D_{s,s})&    
\frac{2}{9}(C_{l,l}- C_{s,s}+2D_{l,l}-D_{l,s}-D_{s,s})\\
0 & \frac{2}{9}(C_{l,l}- C_{s,s}+2D_{l,l}-D_{l,s}-D_{s,s}) 
& \frac{1}{9}(2C_{l,l}+C_{s,s} +4D_{l,l} +4D_{l,s}+D_{s,s}) \\
\end{array}
\right),
\label{eq:suscmatrix}
\end{equation}
where the rows and columns are now labeled by $I$, $Y$ and $B$ in
that order~\cite{NORM}. Note that in the $N_f=3$ simulations for which
$m_u=m_d=m_s$, $C_{l,l}=C_{l,s}=C_{s,s}$ and $D_{l,l}=D_{l,s}=D_{s,s}$,
so $\chi$ is a diagonal matrix in this representation. There are then
no correlations between fluctuations $I$, $Y$ and $B$.
In the $N_f=2+1$ case the only
correlations are between hypercharge and baryon number. Of course
for $m_u\neq m_d$ there would be correlations among the fluctuations
in all three charges.
For temperatures below the phase transition or crossover, the lightest 
particle that can be excited by a chemical potential coupled to the z-component
of isospin is the pion, while for hypercharge and baryon number
chemical potentials it is the kaon and the nucleon, respectively.
Above the transition temperature each of the chemical potentials
can excite quark states that are much lighter than hadrons, so
we expect the diagonal elements of $\chi$ to increase sharply in
the vicinity of the transition, and they do. 

As in the case of $\chi_{\rm tot}$, these susceptibilities can be written as
expectation values of traces of the quark matrices and their derivatives
with respect to the chemical potential, and unbiased estimators of the traces
can be expressed in terms of vectors of Gaussian random numbers~\cite{SUSC}. 
Here too, we use 100 random vectors for each gauge configuration.

\begin{figure}
\centerline{\includegraphics[width=6.0in]{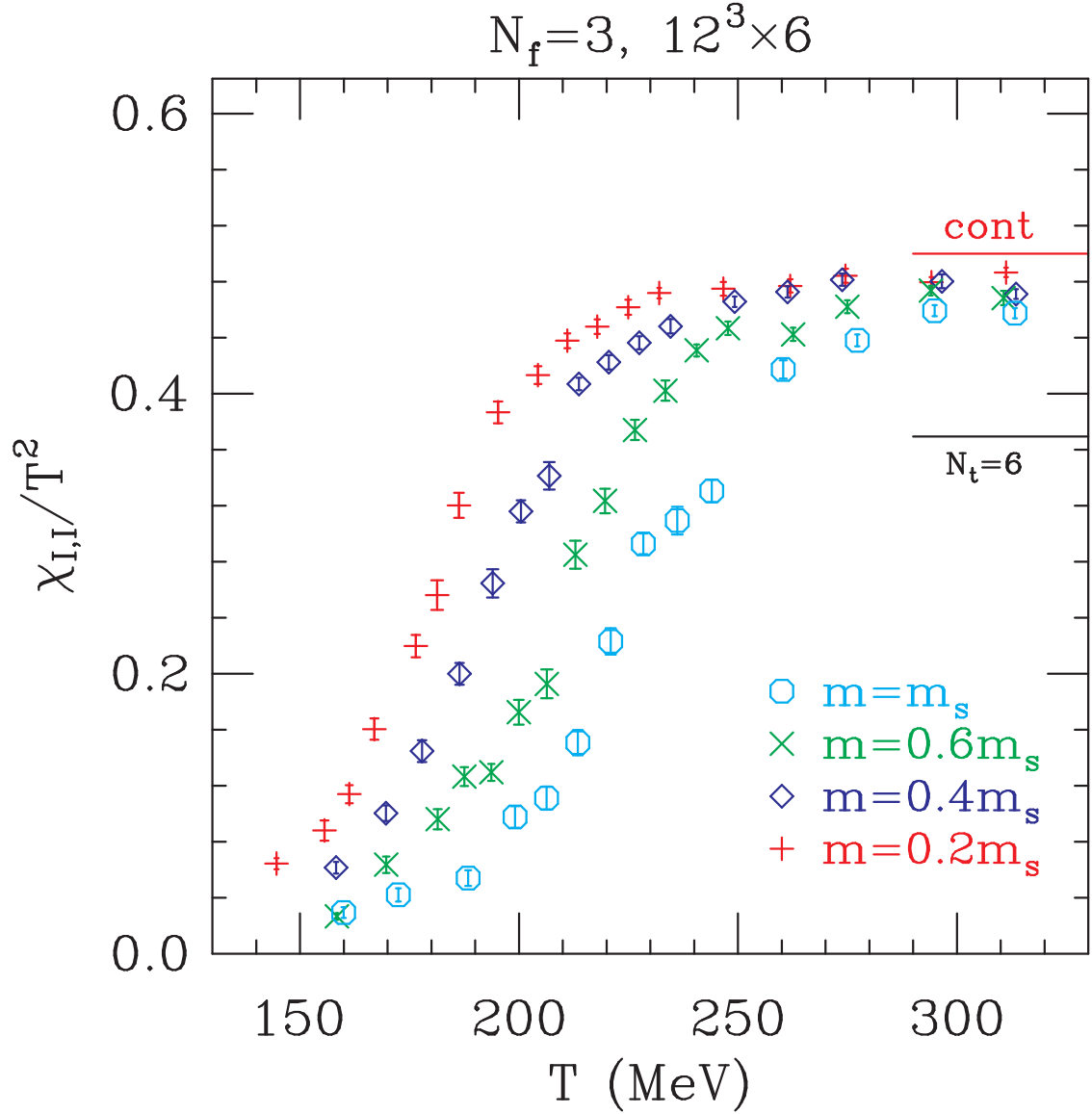}}
\caption{The z-component of isospin susceptibility $\chi_{I,I}$ as
a function of temperature for
three degenerate flavors of quarks with masses $m_q/m_s=1.0$, 0.6, 0.4, 
and 0.2 on $12^3\times 6$ lattices. The solid lines 
on the right of the figure indicate the free quark value in the continuum 
and on a $12^3\times 6$ lattice.
\label{fig:qno_trip_nf3_nt6}
}
\end{figure}

\begin{figure}
\centerline{\includegraphics[width=6.0in]{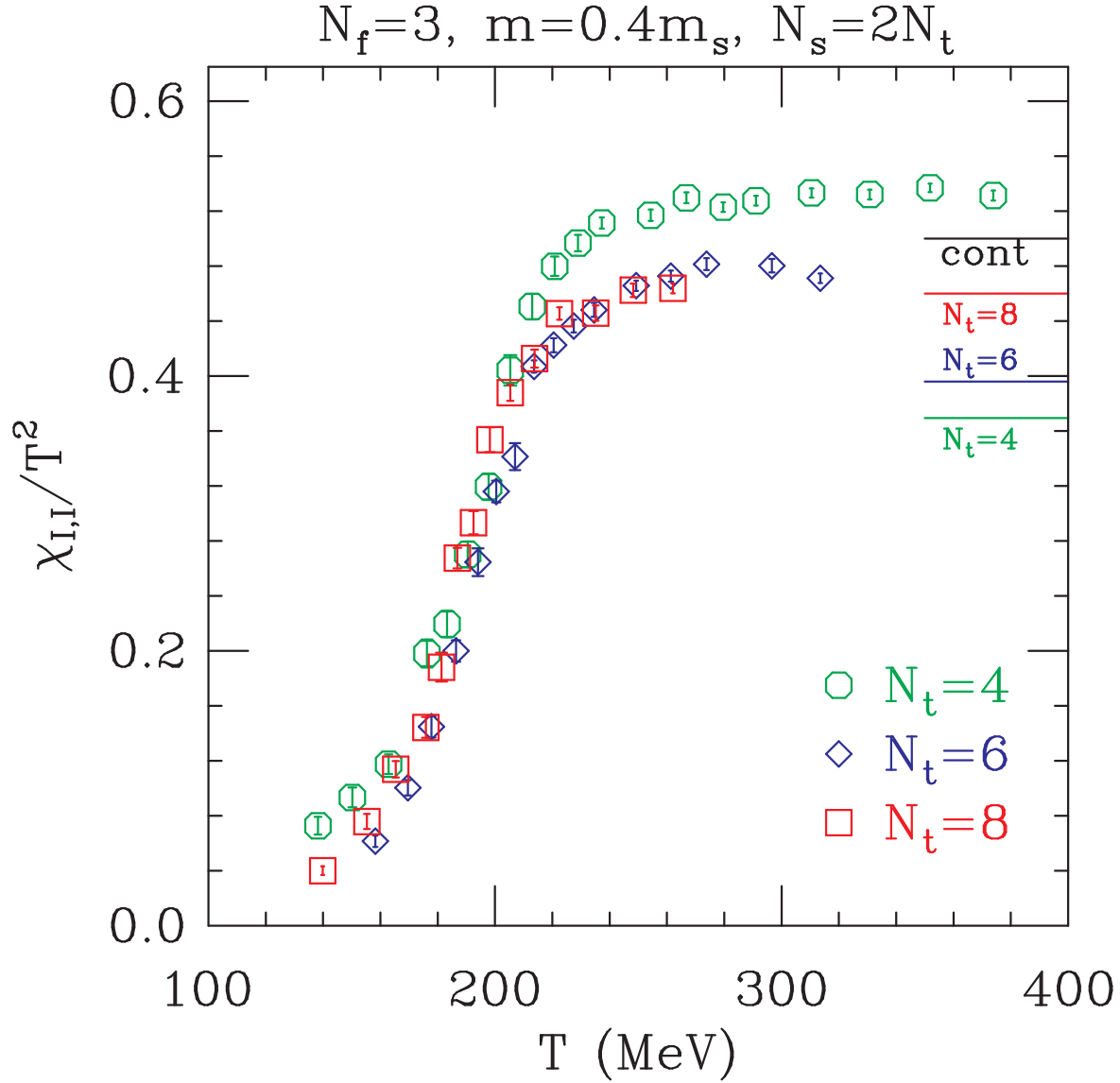}}
\caption{The z-component of isospin susceptibility $\chi_{I,I}$ for
three degenerate flavors of quarks with mass $m_q=0.4\, m_s$, on $8^3\times 4$,
$12^3\times 6$ and $16^3\times 8$ lattices. The solid lines on the right of the
figure indicate the value for free quarks in the continuum, and on the 
finite lattices on which the simulations were carried out. The close 
agreement among these results is another indication
of the excellent scaling properties of the Asqtad action.
\label{fig:qno_trip_nf3_m04}
}
\end{figure}

In Fig.~\ref{fig:EPQFIG} we plot the quark number susceptibility of
massless free quarks for the standard Kogut-Susskind, Wilson, P4 and Asqtad 
actions. (With only one flavor of quark, there is, of course, only
one susceptibility).  One sees that as in the case of the energy and 
pressure, the susceptibility of the Asqtad action is significantly 
closer to the continuum result for small values of $N_t$ (large values 
of the lattice spacing) than that of the Kogut-Susskind
action.  In Fig.~\ref{fig:qno_trip_nf3_nt6} we show $\chi_{I,I}$
for the four quark masses we have studied on $12^3\times 6$ 
lattices with $N_f=3$. One again sees the steepening of the crossover and 
its shift to lower temperature as 
the quark mass is decreased.  In Fig.~\ref{fig:qno_trip_nf3_m04} we 
illustrate the dependence of $\chi_{I,I}$
on lattice spacing by plotting results at fixed quark mass for three
values of $N_t$.  The solid lines on the right of this figure indicate 
the values of the susceptibility for free quarks in the continuum 
and on the finite lattices on which the simulations were carried out.

\begin{figure}
\centerline{\includegraphics[width=6.0in]{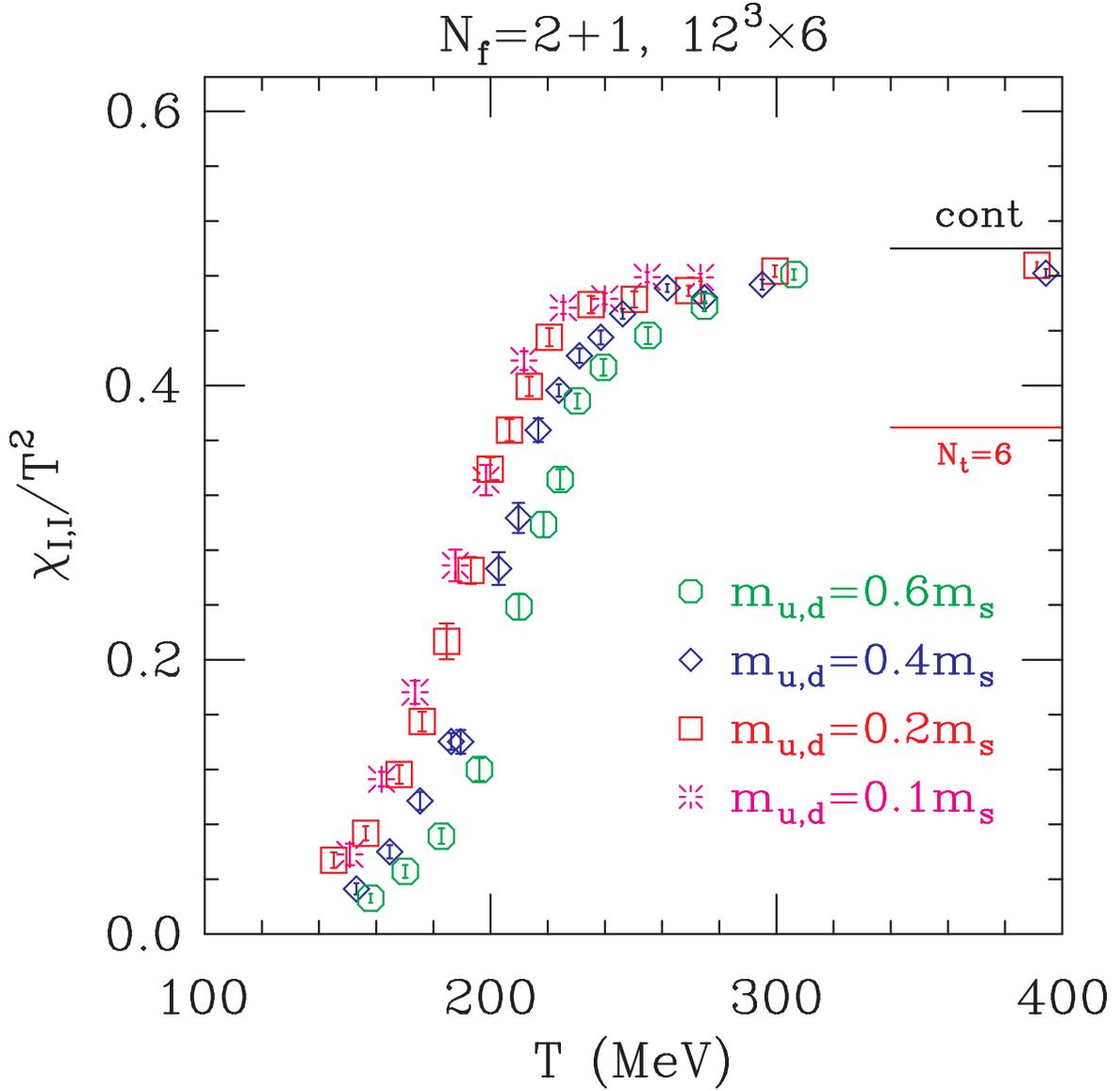}}
\caption{The z-component of isospin susceptibility 
as a function of temperature
for two light and one heavy quark on $12^3\times 6$ lattices. Results are shown
for light quark masses $m_{u,d}/m_s=0.6$, 0.4, 0.2 and 0.1. The
mass of the heavy quark is fixed at $m_s$.  The solid lines on the right of
the figure indicate the free quark value in the continuum and on 
a $12^3\times 6$ lattice.
\label{fig:qno_trip_nf21_nt6}
}
\end{figure}

\begin{figure}
\centerline{\includegraphics[width=6.0in]{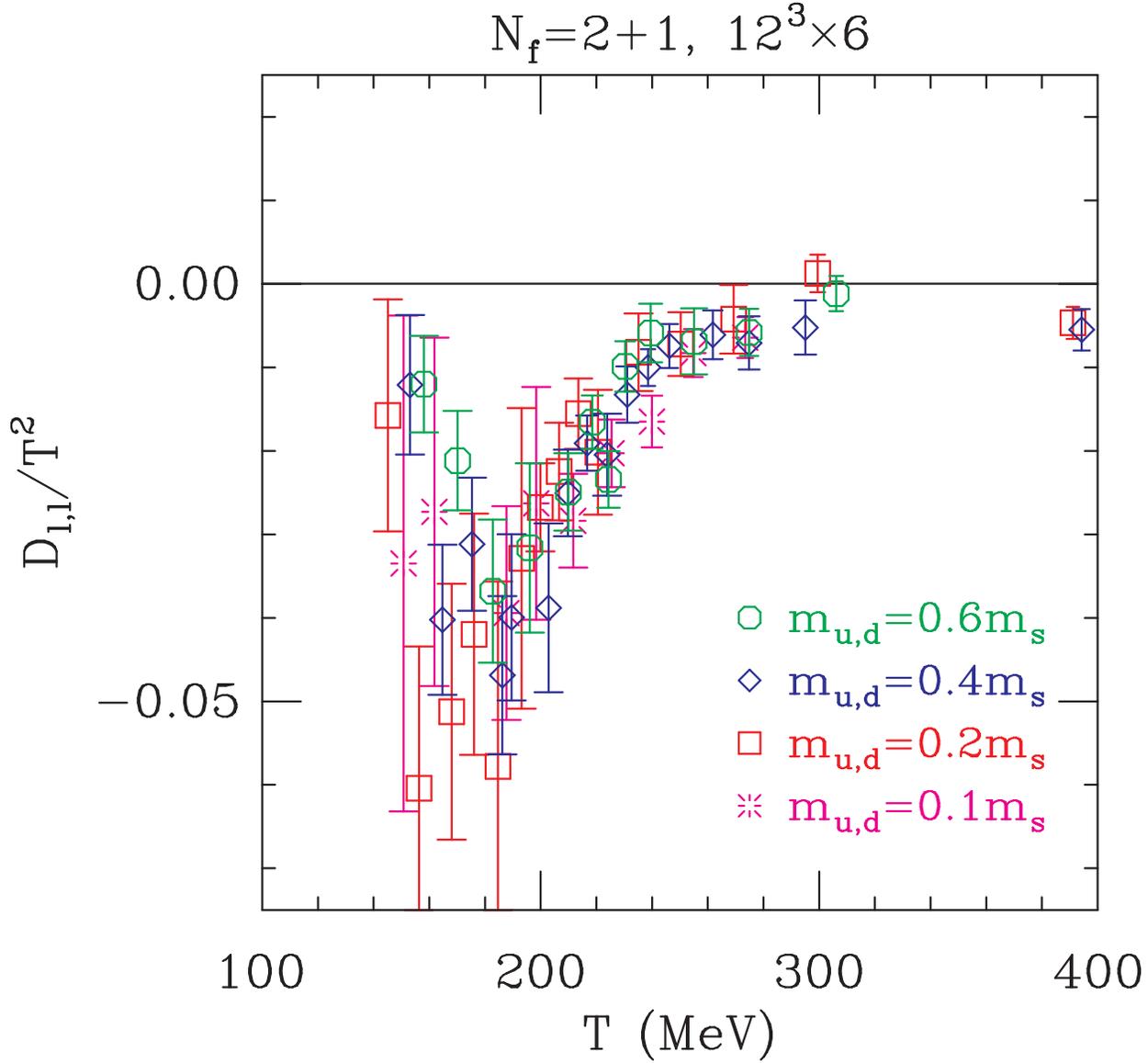}}
\caption{The disconnected part of the light quark susceptibility 
$D_{l,l}$ as a function of temperature for two light and one heavy 
quark on $12^3\times 6$ lattices. Results are shown for light quark 
masses $m_{u,d}/m_s=0.6$, 0.4, 0.2 and 0.1. The
mass of the heavy quark is fixed at $m_s$.
\label{fig:qno_diff_nf21_nt6}
}
\end{figure}

\begin{figure}
\centerline{\includegraphics[width=6.0in]{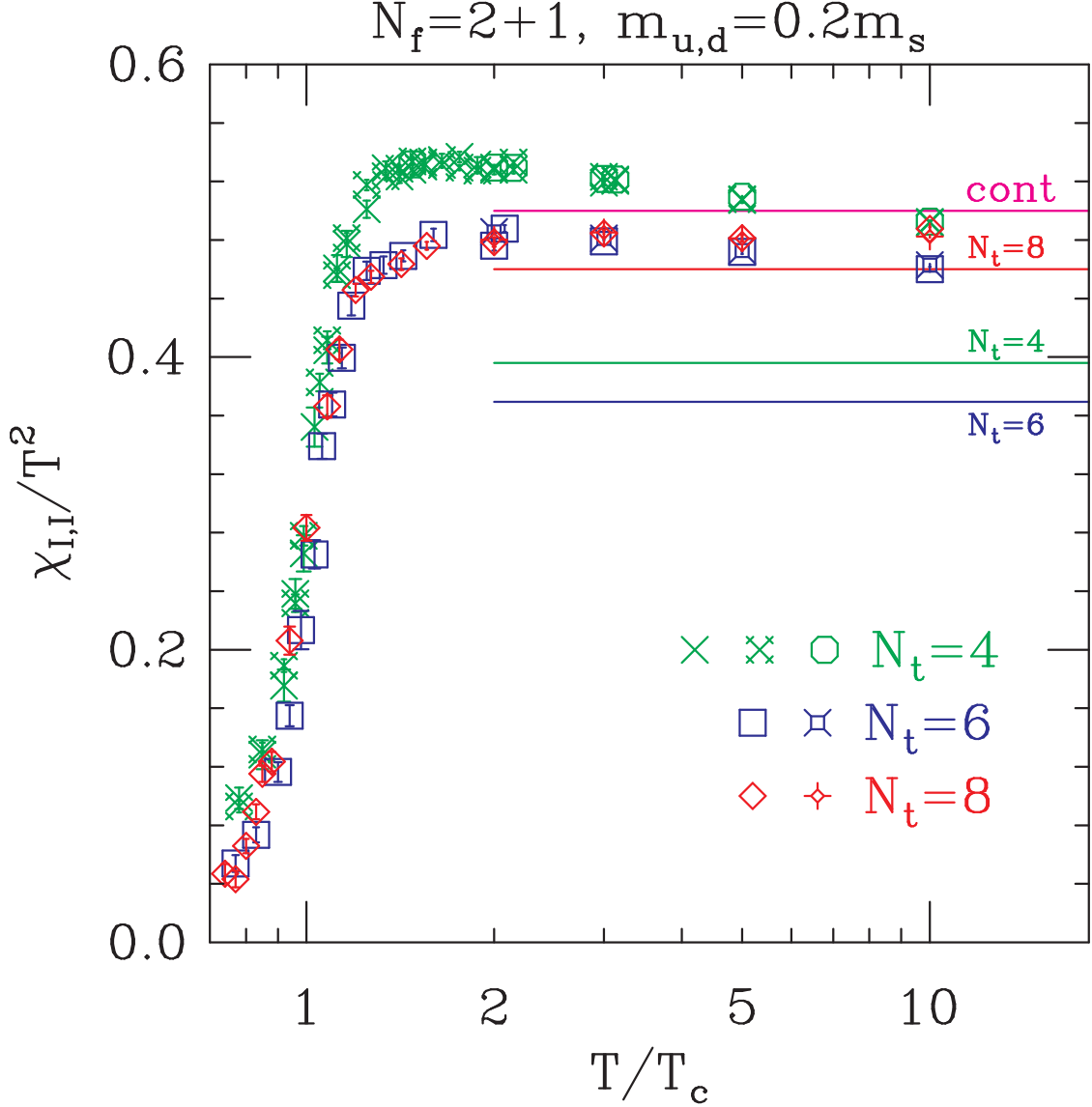}}
\caption{The z-component of isospin susceptibilities as a 
function of $T/T_c$ 
for two light quarks with mass $0.2\, m_s$ and one heavy quark with mass $m_s$.
Results are shown for lattices with 4, 6 and 8 time slices. For $N_t=4$
the octagons, fancy crosses, and crosses are data from spatial volumes
$16^4$, $12^3$ and $8^3$, respectively. For $N_t=6$ the fancy squares
and squares are data from spatial volumes $18^3$ and $12^3$, while for
$N_t=8$ the fancy diamonds and diamonds are data from spatial volumes
$24^3$ and $16^3$.  The solid lines on the right of the figure 
indicate the value for free quarks in the continuum and on the finite 
lattices on which the simulations were carried out. The agreement between 
the $N_t=6$ and 8 results illustrate the excellent scaling property 
of the Asqtad action, and indicates that these results are close to the
continuum ones.
\label{fig:qno_trip_nf21_0.2ms}
}
\end{figure}

\begin{figure}
\centerline{\includegraphics[width=6.0in]{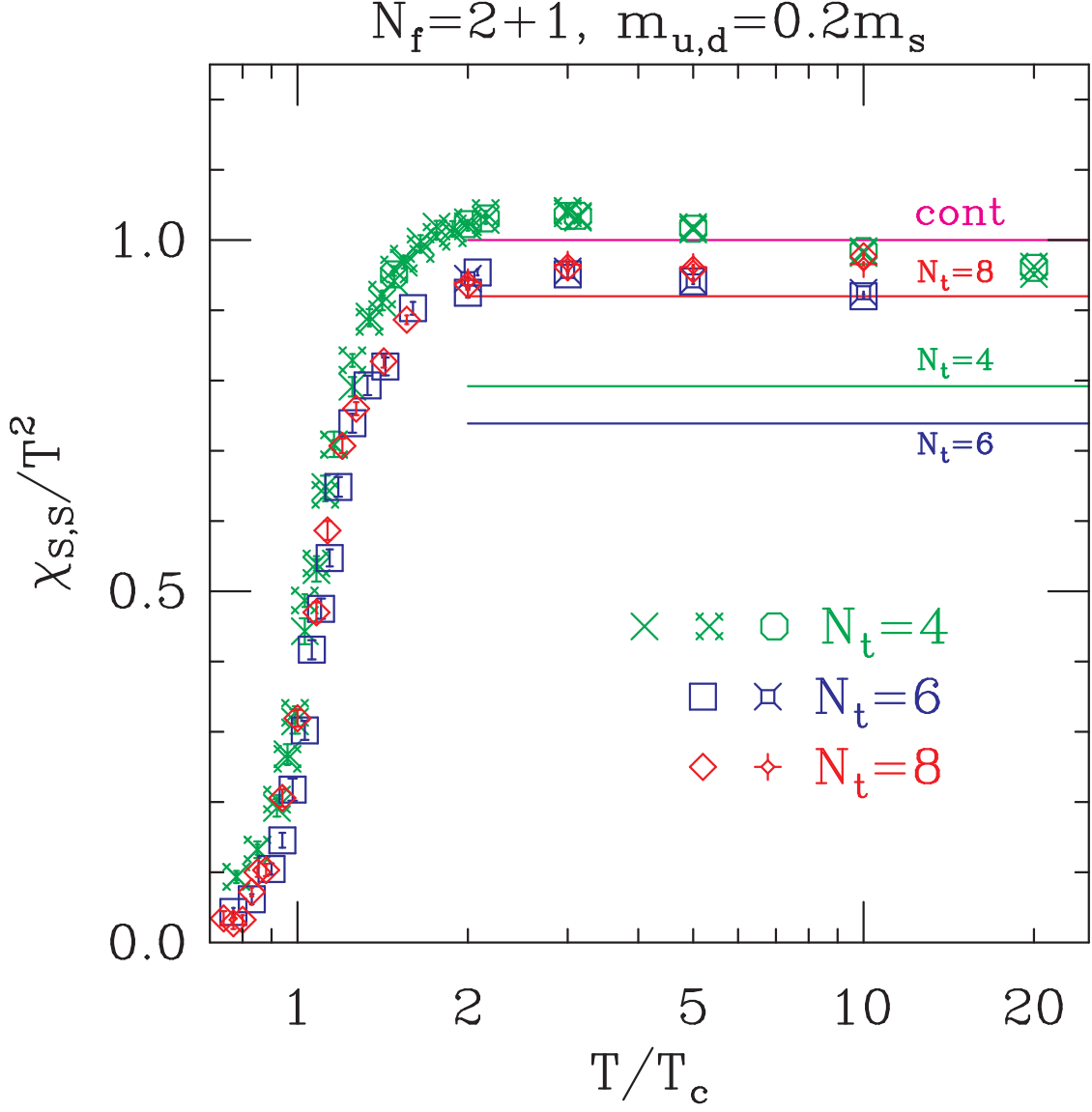}}
\caption{The strange quark number susceptibility $\chi_{S,S}$
as a function of $T/T_c$
for two light quarks with mass $0.2\, m_s$ and one heavy quark with mass $m_s$.
Results are shown for lattices with 4, 6 and 8 times slices. The symbols
have the same significance as in Fig.~\ref{fig:qno_trip_nf21_0.2ms}. The solid 
lines on the right of the figure indicate the value for free quarks in the 
continuum and on the finite lattices on which the simulations were carried out. 
\label{fig:qno_strange}
}
\end{figure}

\begin{figure}
\centerline{\includegraphics[width=6.0in]{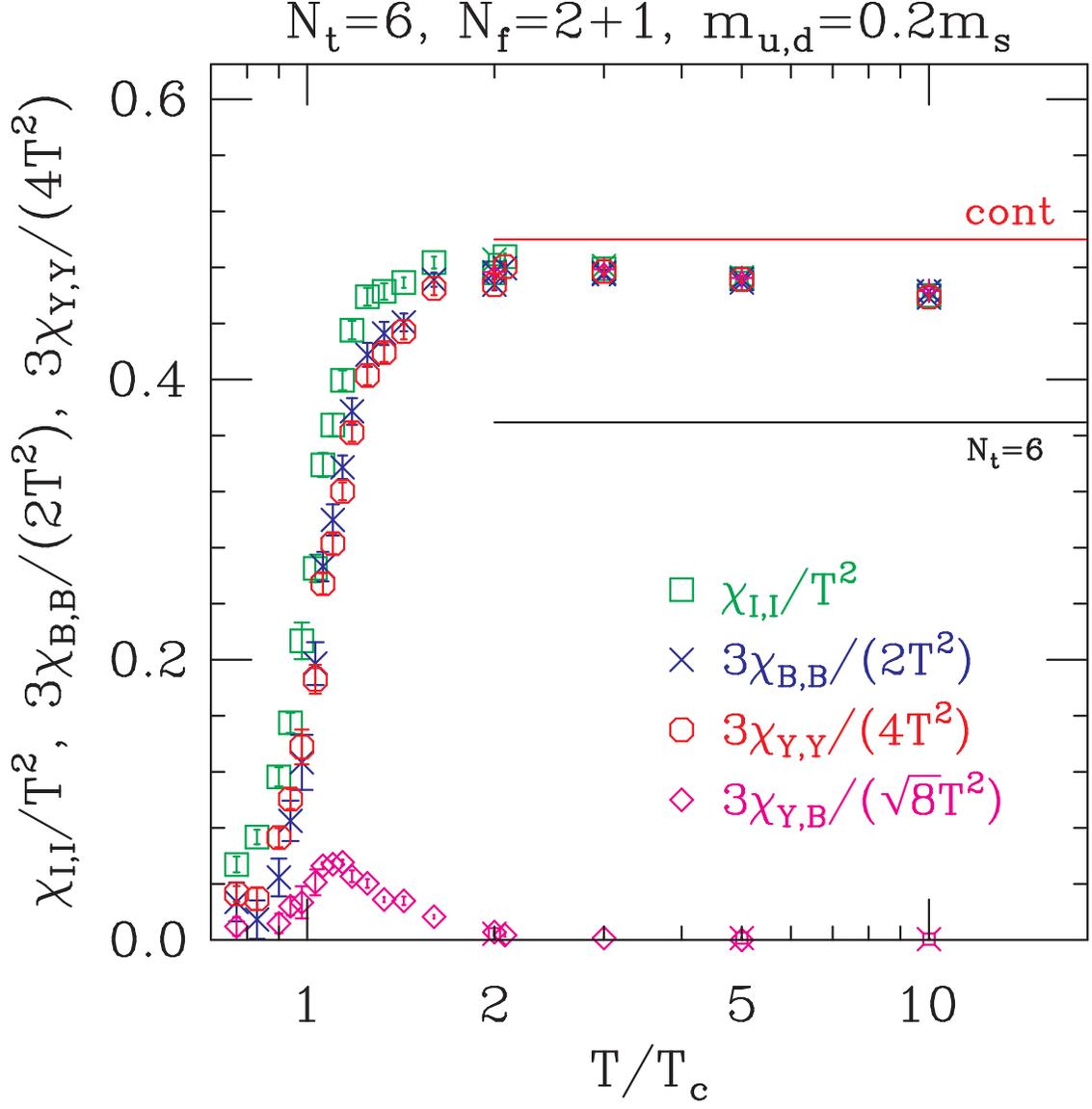}}
\caption{The diagonal elements of the susceptibility matrix,
$\chi_{I,I}$, $\chi_{B,B}$ and $\chi_{Y,Y}$ as a function of
$T/T_c$ for two light quarks with mass $0.2\, m_s$ and one 
heavy quark with mass $m_s$ on $12^3\times 6$ lattices.
$\chi_{Y,Y}$ and $\chi_{B,B}$ have been normalized so that they
approach the same limit as $\chi_{I,I}$ at high temperatures. Also shown 
is the off diagonal matrix element $\chi_{Y,B}$, which measures
correlations between fluctuations in the hypercharge and baryon
number. The coefficient of $\chi_{Y,B}$ is the geometric
mean of those for $\chi_{Y,Y}$ and $\chi_{B,B}$.
\label{fig:qno_combo_new}
}
\end{figure}

In Fig.~\ref{fig:qno_trip_nf21_nt6} we plot $\chi_{I,I}$
on $12^3\times 6$ lattices for $N_f=2+1$. As indicated in
Eq.~(\ref{eq:suscmatrix}), this quantity is proportional
to the light quark connected diagram, $C_{l,l}$. The
disconnected diagrams are, of course noisier. As an example,
we plot the light quark disconnected graph in
Fig.~\ref{fig:qno_diff_nf21_nt6}. This quantity is given by

\begin{equation}
D_{l,l} = \langle (B+Y/2)^2\rangle 
-\langle(B+Y/2)\rangle^2
         - \chi_{I,I}.
\end{equation}
We see in Fig.~\ref{fig:qno_diff_nf21_nt6} that $D_{l,l}$
can be cleanly evaluated in the neighborhood of the crossover, 
the only region in which it is appreciable. 
Below the crossover one expects $\chi_{I,I}$ to be larger than 
$\chi_{B+Y/2,B+Y/2}$, since in this regime the lowest
energy state that can be excited by a chemical potential coupled to
$I_z$ is a pion, while the lowest energy state that can be
excited by a chemical potential coupled to $B+Y/2$ 
is a kaon.  Our results suggest that vestiges of hadronic 
physics persist in the plasma at least up to 240~MeV.
In Figs.~\ref{fig:qno_trip_nf21_0.2ms} and \ref{fig:qno_strange}
we show results for $\chi_{I,I}$ and for the strange quark number 
susceptibility, $C_{s,s}+D_{s,s}$, on lattices with 4, 6 and 8 time
slices for a range of spatial volumes. There is no observable 
dependence on the spatial volume. The close agreement between the $N_t=6$ 
and 8 results here and for $N_f=3$ illustrates the excellent scaling 
properties of the action, and indicates that our results are close 
to the continuum ones. Finally, in Fig.~\ref{fig:qno_combo_new}
we plot the diagonal elements of the susceptibility matrix,
$\chi_{I,I}$, $\chi_{Y,Y}$ and $\chi_{B,B}$, as a function of
temperature for two light quarks with mass $0.2\, m_s$ and one 
heavy quark with mass $m_s$ on $12^3\times 6$ lattices. $\chi_{Y,Y}$
and $\chi_{B,B}$ have been multiplied by factors of 3/4 and 3/2
respectively, so that the quantities plotted approach the same high 
temperature limit as $\chi_{I,I}$. Also shown is $\chi_{Y,B}$,
the only non-zero off-diagonal matrix element of $\chi$ for
$m_u=m_d$. It measures correlations between fluctuations in the 
hypercharge and baryon number. The coefficient of $\chi_{I,B}$ in this
figure is the geometric mean of those for $\chi_{Y,Y}$ and $\chi_{B,B}$.

\section{CONCLUSION}

For the quark masses we have studied to date, we find a rapid crossover,
rather than a {\it bona fide} phase transition, for both $N_f=3$ and
$N_f=2+1$. Our result for $N_f=3$ is consistent with recent work of Karsch 
{\it et al.}~\cite{P4MASS}. They find a first order phase transition for
$N_f=3$ only for pion masses below 290(20)~MeV for the standard Kogut-Susskind
action and below 67(18)~MeV for the improved P4 action on lattices with four
time slices. The lightest quark mass used in~\cite{P4MASS} corresponded to
a pion mass of 170~MeV, and reweighting techniques were used to extrapolate to
lighter quark masses. Our lightest $N_f=3$ quark mass corresponds to a pion mass of
approximately 340~MeV. Because of the large difference between standard and 
improved actions on these lattices, it seems particularly important to push our work
at $N_t=6$ and 8 to smaller quark masses, and we intend to do so. 

The small quark mass that seems to be required for a first order transition 
at $N_f=3$ strongly suggests that in the real world, $N_f=2+1$, there is no phase
transition at the physical quark masses. With this assumption, we have estimated
the critical temperature for $N_f=2+1$ at $m_{u,d}=0$ through an extrapolation
of the form

\begin{equation}
r_1\,T_c=c_0+c_1\,(m_\pi/m_\rho)^d + c_2\, (aT_c)^2,
\end{equation}

\noindent
where we evaluated $T_c$ for each value of $N_t$ and $m_\pi/m_\rho$ for which
we have made measurements from the peak in the $\bar\psi\psi$ susceptibility.
For a second order phase transition in the O(4) universality class at
$m_{u,d}=0$, $d=2/\beta\delta\approx 1.08$.
We find that  $T_c=169(12)(4)$~MeV with a $\chi^2$ of 2.1 for 11 degrees
of freedom. The first error is the fit error, the second from the uncertainty
in $r_1$, taken as 0.317(7)~fm~\cite{NEWSPEC}.
To test the sensitivity of $T_c$ to $d$, we have also performed
a fit with $d=2$, which yields $T_c=174(11)(4)$~MeV with a 
$\chi^2$ of 1.5 for 11 degrees of freedom. 
So, the goodness of the fit does not allow us to prefer either of them.

Finally, we note that the agreement between the quark number susceptibilities
on $N_t=6$ and 8 lattices is very encouraging, as it indicates that our
results for these quantities are close to their continuum values.

\section*{ACKNOWLEDGMENTS}
Computations for this work were performed at Florida State
University, Fermi National Accelerator Laboratory (FNAL), the National 
Center for Supercomputer Applications (NCSA), 
the National Energy Resources Supercomputer Center (NERSC),
and the University of Utah (CHPC). 
This work was supported by the U.S. Department of Energy under contracts
DE-FG02-91ER-40628,      
DE-FG02-91ER-40661,      
and
DE-FG03-95ER-40906       
and National Science Foundation grants
PHY01-39929              
and
PHY00--98395.            
TB acknowledges current support from BMBF and GSI.

\newpage

\begin{thebibliography}{9}
\bibitem{NORM} Our normalization of the quark number susceptibilities
differs from that used in studies of two-flavor QCD and
in our earlier reports on this work. In our present notation
$\chi_{I,I}=\chi_{NS}/4$ and $\chi_{B+\frac{1}{2}Y, B+\frac{1}{2}Y}=\chi_S/4$,
where $\chi_{NS}$ and $\chi_S$ are commonly referred to as the non-singlet
and singlet susceptibilities.
\end{thebibliography}
\end{document}